\newcommand{\ju}[1]{\textcolor{red}{{\it#1}} }

\documentclass{cimento}
\usepackage{graphics}
\usepackage{graphicx}
\usepackage{amssymb,amsmath}
\usepackage{tabularx,booktabs}

\usepackage{color}

\newcommand{\henry}[1]{\textcolor{red}{#1}}
\title{The Hydrodynamics of Active Systems}

\author{Julia M. Yeomans}
\instlist{\inst{The Rudolf Peierls Centre for Theoretical Physics, 1 Keble Road, Oxford, OX1 3NP, UK}}

\begin{document}
\maketitle

\begin{abstract}

This is a series of four lectures presented at the 2015 Enrico Fermi summer school in Varenna. The aim of the lectures is to give an introduction to 
the hydrodynamics of active matter concentrating on low Reynolds number examples such as cells and molecular motors. 
Lecture 1 introduces the hydrodynamics of single active particles, covering the Stokes equation and the Scallop Theorem, and stressing the link between autonomous activity and the dipolar symmetry of the  far flow field.
In lecture 2 I discuss applications of this mathematics to the behaviour of microswimmers at surfaces and in external flows, and describe our current understanding of how swimmers stir the surrounding fluid. 
Lectures 3  concentrates on the collective behaviour of active particles, modelled as an active nematic.  I write down the equations of motion and motivate the form of the active stress. The resulting hydrodynamic instability leads to a state termed `active turbulence' characterised by strong jets and vortices in the flow field and  the continual creation and annihilation of pairs of topological defects.
Lecture 4 compares simulations of active turbulence to experiments on suspensions of microtubules and molecular motors. I introduce lyotropic active nematics and discuss active anchoring at interfaces.

\end{abstract}

\section{Introduction}

Active systems produce energy at the level of their individual components and naturally exist out of thermodynamic equilibrium. The most obvious examples are living matter, animals, plants and bacteria, but inanimate systems, such as vertically vibrated granular layers, are also considered active, and there are many parallels between the behaviour of active and driven matter. Active systems occur across length scales. Well studied macroscopic examples include flocks of birds and schools of fish. On a micron length scale, swimming bacteria and crawling and dividing cells provide a rich testing ground for the theories of active matter. Within cells there is a highly active environment, with molecular motors of size a few nanometres converting chemical to mechanical energy to organise cellular processes and hence power life.

Why is active matter of interest to a physicist? We are dealing with energy, work and motion, and often with fluctuations, standard physical concepts. Describing the motion of swimming bacteria has an august place in the history of fluid mechanics. The collective behaviour of active systems appears to have many similarities across length scales, suggesting that understanding based on generic principles rather than individual details might be useful. Moreover experiments on active matter provide a testing ground for the theories of non-equilibrium statistical physics.

Scientists are also taking the first steps towards an active technology. Active colloids are autonomous micro-machines and miniature driven swimmers can carry payloads and mix fluids in microscopic channels. Bacteria are able to turn micron-scale flywheels and to sort passive colloids. Even more exciting is the future possibility of imitating cellular transport and organisation, processes which have evolved over millennia to maximise efficiency in the strongly fluctuating environment of the cell.

These lectures concentrate on the hydrodynamics of active matter; the behaviour in a background fluid. I shall first consider single microswimmers, such as a swimming bacterium. Because of their size, the motion of microswimmers is governed by the Stokes equations, the low Reynolds number limit of the Navier-Stokes equations. This has two far-reaching consequences, the Scallop Theorem which controls the form of the swimming stroke, and the nematic symmetry of the swimmers' far flow field. After explaining these concepts and discussing the mathematics behind them I shall show how they impact upon the interaction between swimmers and tracer particles, and how microswimmers move near a surface and in Poiseuille flow.

The second half of the lectures will deal with the hydrodynamics of dense suspensions of active nematic particles. I shall show how a simple extension of the continuum equations of motion of nematic liquid crystals describes active hydrodynamics and leads to a state termed active turbulence characterised by a continually changing pattern of jets and vortices in the velocity field and by the creation and annihilations of walls and topological defects in the director field. Active turbulence has been observed in dense swimmer suspensions, in confluent cellular layers and in mixtures of microtubules and molecular motors. I shall compare experiments with numerical solutions of the active nematohydrodynamic equations and discuss the mechanisms behind active turbulence. 

\section{Single swimmer hydrodynamics: background}

Lecture 1 describes the hydrodynamics of a single microswimmer and starts by introducing the Stokes equations as the low Reynolds number limit of the Navier-Stokes equations. I motivate the Scallop Theorem, that a swimming stroke must be non-invariant in time for a net swimmer motion, and show how this relates to bacterial swimming strategies. I then introduce the multipole expansion and argue that, because swimmers are force free, their far flow field is generically dipolar, with nematic symmetry  \cite{LP09,EW15}.

\subsection{Swimming at low Reynolds number}
I write down the continuum equations of motion of a simple fluid and take the zero Reynolds number limit, appropriate to many colloidal and active systems, to obtain the Stokes equations.\\
~\\
The Navier-Stokes equations, that provide a continuum description of the flow of a Newtonian, incompressible fluid, are  \cite{LL00}
\begin{equation}
\rho \left\{\frac{\partial \mathbf{v}}{\partial t} +(\mathbf{v} \cdot \nabla ) \mathbf v \right \}
= - \nabla p + \mu \nabla^2 \mathbf{v} + \mathbf{f}, \quad \quad  \nabla \cdot {\mathbf{v}}=0,
\label{eq:NS}
\end{equation}
where $\mathbf v (\mathbf{r},t)$ is the velocity at position $\mathbf{r}$ and time $t$ of a fluid of density $\rho$ and dynamic viscosity $\mu$ driven by a pressure gradient $\nabla p$ and a body force (force per unit volume) $ \mathbf{f}$. The terms on the left hand side of Eq.~(\ref{eq:NS}) are the inertial terms which describe the transport of momentum and $\mu \nabla^2 \mathbf{v}$ describes the viscous dissipation that results from velocity gradients.

Dimensionless variables denoted by a tilde, can be defined by choosing a length scale $L_0$ and a velocity scale $V_0$ 
\begin{equation}
\tilde{v}=\frac{v}{V_0},\;\;  \quad \quad \quad \tilde{x}=\frac{x}{L_0}, \quad \quad \quad
\tilde{\nabla}=L_0 \nabla, \quad \quad \quad\tilde{t}=\frac{V_0}{L_0}t,\quad \quad\quad  \frac{\partial}{\partial \tilde{t}}= \frac{L_0}{V_0} \frac {\partial}{\partial t}.
\end{equation}
In terms of the dimensionless variables the Navier-Stokes equation becomes
\begin{equation}
\left\{\frac{\partial \mathbf{\tilde{v}}}{\partial \tilde{ t}} +(\mathbf{\tilde{v}} \cdot \tilde{\nabla} ) \mathbf {\tilde{ v}} \right \}
= - \frac{L_0}{V_0^2 \rho} \nabla p + \frac{\mu}{L_0 V_0 \rho} \tilde{\nabla}^2 \mathbf{\tilde{v}} + \frac{L_0}{V_0^2 \rho} \mathbf{f}.
\label{eq:scaledNS}
\end{equation}
Eq.~(\ref{eq:scaledNS}) shows that relative magnitude of the inertial and viscous terms in the Navier-Stokes equation is characterised by a dimensionless number, the Reynolds number
 \begin{equation}
\mathrm{Re}=\frac{\mathrm{inertial \;response}}{\mathrm{viscous\;\; response}} \sim \frac{\rho L_0 V_0}{\mu}.
\end{equation}
For water $\rho/\mu \sim 10^{6} $, so colloids moving in microflows and micro-swimmers with length scales $\sim 1-100 \mu$m and velocity scales $\sim 1-100 \mu$ms$^{-1}$ have Reynolds numbers $\ll 1$. Thus the inertial terms can be neglected, and the Navier-Stokes equations reduce to the Stokes equations
\begin{equation}
\nabla p = \mu \nabla^2 \mathbf{v} + \mathbf{f},\quad \quad \quad\quad  \nabla \cdot \mathbf{v}=0.
\label{eq:Stokes}
\end{equation}
The zero $\mathrm{Re}$ limit has interesting physical consequences. Moreover the Stokes equations are linear and hence more tractable mathematically.


\subsection{The Scallop Theorem}

\begin{figure}[htdp]
\includegraphics[trim = 0 0 0 0, clip, width = \linewidth]{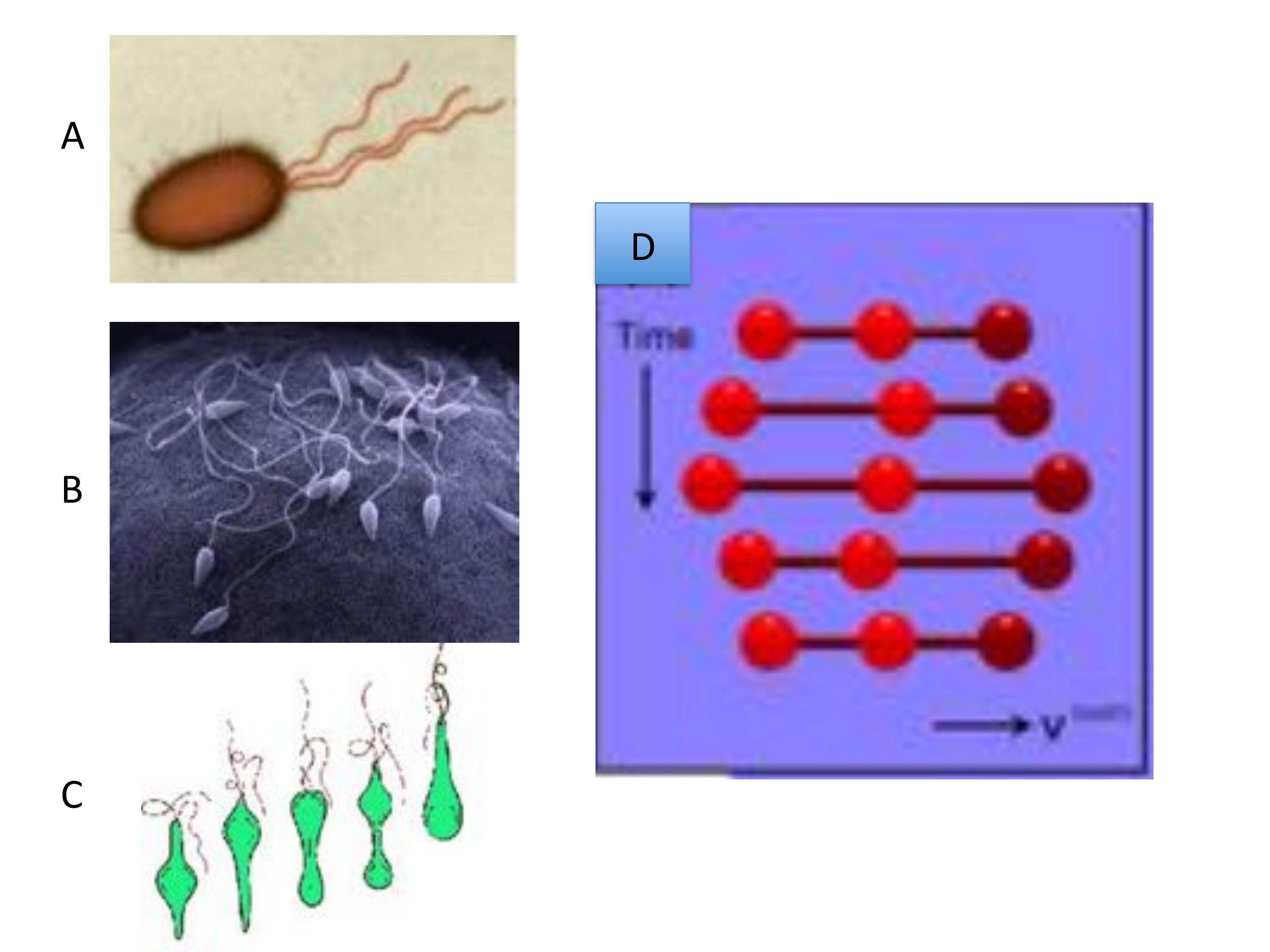}
\caption{Microscopic swimmers that have evolved to overcome the Scallop Theorem. A: {\it E. coli}, B: sperm cells, C: {\it Euglena metaboly}, D: the 3-sphere swimmer (after Najafi and Golestanian  \cite{NG04}).}
\label{swimmers}
\end{figure}

A far-reaching consequence of the $\mathrm{Re=0}$ limit is Purcell's Scallop Theorem: To achieve propulsion at zero Reynolds number in Newtonian fluids a swimmer must deform in a way that is not invariant under time reversal.\\
~\\
There is no explicit time dependence in the Stokes equations which implies kinematic reversibility. A striking example of this is provided by a blob of dye suspended in a high viscosity fluid contained between two cylinders. When the fluid is sheared by rotating the outer cylinder the dyed blob can be stretched to wrap several times around the inner cylinder. If the motion of the cylinders is reversed the dyed fluid returns to its original shape.

This is the physics behind the Scallop Theorem  \cite{P77}. As a swimmer moves through a low $\mathrm{Re}$ fluid there is nothing in the equations of motion of the surrounding fluid that picks out a preferred direction. If the boundary conditions, i.e.\ the swimmer's stroke cycle, is invariant under time reversal as well, its net displacement after each cycle must be zero -- otherwise reversing time would give the same physical system, but a different displacement. Therefore at zero $\mathrm{Re}$ the stroke must be non-invariant under time reversal to allow swimming.

Microscopic swimmers have evolved several strategies to overcome the Scallop Theorem (Fig.~1). Many move by using long, thin  appendages called flagella. Bacterial flagella (e.g.\ {\it E. coli}) are driven by rotary motors which induce helical waves to overcome the Scallop Theorem. Eukaryotic flagella tend to oscillate through bending waves. {\it Paramecium} is covered by cilia, shorter but similar in structure to eukaryotic flagella that have distinct power and recovery strokes. Cilia also act as pumps in the body, clearing mucus from the lungs and in the fallopian tubes to facilitate moving the ovum from the ovary to the uterus. {\it Euglena metaboly} produces a suitable swimming stroke by altering its body shape so that the relative motion of its ends is coupled to changes in their mass.

At least three degrees of freedom are needed to define a suitable stoke for zero $\mathrm{Re}$ swimming. A simple model swimmer is the three-sphere swimmer, which comprises three spheres coupled by rods  \cite{NG04}. The rods have no hydrodynamic coupling to the fluid, but specifying their lengths as a function of time serves to define the swimming stroke. The stroke and the swimmer displacement as a function of time are shown in Fig.~1. The advantage of building the swimmer out of spheres is that the flow field around a sphere is known analytically in the Stokes limit  \cite{LL00}. Hence exact results can be obtained for the swimming speed if the sphere radii are small compared to their separation.

Another widely used model swimmer is the squirmer  \cite{L52,B71}, useful because it admits an exact solution and also because it is spherical which helps to disentangle steric and hydrodynamic effects in simulations. The squirmer is a sphere of radius $a$, and motion is imposed by 
choosing suitable velocity boundary conditions on the surface of the sphere.  Defining a spherical polar co-ordinate system with the squirmer velocity along $z$, the radial velocity at the surface is taken to be zero and the tangential surface velocity is a power series in the first derivatives of the Legendre polynomials
\begin{equation}
v_{\theta}(a,\theta)=\sum_{n=1}^{\infty} B_n \frac{2}{n(n+1)}\sin{\theta}P_n^{\prime}(\cos{\theta}),
\end{equation};l..
where the $B_n$ can depend on time.

Consider the simplest non-trivial example, $B_n=0, \quad n>2$. In the rest frame of the squirmer, which is moving with velocity $\frac{2}{3}B_1$ along the $z$-axis, it is easy check that the solution to the Stokes equations that obeys these boundary conditions is
\begin{eqnarray}
v_r(r,\theta)& =& \frac{2}{3} B_1 \cos{\theta}  - \frac{a^2}{r^2}\frac{ B_2}{2} (3\cos^2{\theta}-1) + \frac{2 B_1}{3} \frac{a^3}{r^3} \cos{\theta} + O\left(\frac{a^4}{r^4}\right) ,     \nonumber\\
v_{\theta}(r,\theta)& =&- \frac{2}{3} B_1 \sin{\theta}+ \frac{B_1}{3} \frac{a^3}{r^3} \sin{\theta}   + O\left(\frac{a^4}{r^4}\right).
\label{eq:squirmer}
\end{eqnarray}
Note that there is no term in $r^{-1}$ in the far field expansion~(\ref{eq:squirmer}). This is a general feature of active systems and in the next section we explain why.


\subsection{Far flow fields}
We introduce the multipole expansion, a far field expansion for the flow field around a localised force distribution. For active systems, with no applied forces or torques, the far field velocity has dipolar symmetry.\\
~\\
For a point force $\mathbf{f}$ acting at the origin the Stokes equations~(\ref{eq:Stokes}) can be solved exactly. The resulting velocity field, termed the Stokeslet, and the corresponding pressure field, at a relative position $\mathbf{r}$ from the swimmer are  \cite{P92,KK05}
\begin{equation}
\mathbf{v}=\frac{\mathbf{f}}{8 \pi \mu} \cdot \left( \frac{\mathbf{I}}{r} + \frac{\mathbf{r}\mathbf{r}}{r^3} \right) \equiv \mathbf{G} \cdot \mathbf{f}, 
\quad\quad\quad
p=p_0 + \frac{\mathbf{f}\cdot \mathbf{r}}{4 \pi r^3}.
\label{eq:Stokeslet}
\end{equation}
In Eq.~(\ref{eq:Stokeslet}) $\mathbf{I}$ is the unit tensor, and $p_0$ is a reference, constant pressure. $\mathbf{G}$ is the Greens function of the Stokes equations, often called the Oseen tensor (and often defined having removed the factor $\frac{1}{8 \pi \mu}$).          There are several ways of obtaining the Stokeslet, none of them very simple. A clear and helpful list of the possible approaches is given by Maciej Lisicki   at  http://arxiv.org/abs/1312.6231.

We extend Eq.~(\ref{eq:Stokeslet}) to a force distribution $\mathbf{f}(\mathbf{\xi})$
\begin{equation}
v_i(\mathbf{r}) = \int G_{ij} (\mathbf{r}-\mathbf{\xi}) f_j(\mathbf{\xi}) \;d\mathbf\xi,
\label{firststep}
\end{equation}
where
\begin{equation}
G_{ij}(\mathbf{r}-\mathbf{\xi})=\frac{1}{8 \pi \mu} \left( \frac{\delta_{ij}}{\mid \mathbf{r}-\mathbf{\xi} \mid} + \frac{(\mathbf{r}-\mathbf{\xi})_i (\mathbf{r}-\mathbf{\xi})_j }{\mid \mathbf{r}-\mathbf{\xi} \mid^3} \right).
\end{equation}
Taylor expanding about the origin
\begin{eqnarray}
v_i(\mathbf{r})& = &\int \left\{ G_{ij} (\mathbf{r}) 
-\frac{\partial G_{ij}}{\partial \xi_k} (\mathbf{r})\xi_k
+\frac{1}{2}\frac{\partial^2 G_{ij}}{\partial \xi_k \partial \xi_l} (\mathbf{r})\xi_k \xi_l \ldots
\right  \}
 f_j(\mathbf{\xi}) \;d\mathbf{\xi} \nonumber\\
&=& G_{ij} (\mathbf{r}) \int f_j(\mathbf{\xi}) \;d\mathbf{\xi}
-\frac{\partial G_{ij}}{\partial \xi_k} (\mathbf{r}) \int \xi_k f_j(\mathbf{\xi}) \; d\mathbf{\xi}
+ \frac{1}{2}\frac{\partial^2 G_{ij}}{\partial \xi_k \partial \xi_l} (\mathbf{r})
 \int \xi_k \xi_l  f_j(\mathbf{\xi}) \; d\mathbf{\xi} + \ldots  \nonumber\\
&\equiv&  G_{ij} (\mathbf{r}) F_j +  \frac{\partial G_{ij}}{\partial \xi_k} (\mathbf{r}) D_{jk}
 +\frac{1}{2}   \frac{\partial^2 G_{ij}}{\partial \xi_k \partial \xi_l}   (\mathbf{r}) Q_{jkl} +\ldots
\label{eq:multipole}
\end{eqnarray}
The first term in Eq.~(\ref{eq:multipole}), $G_{ij} (\mathbf{r}) F_j$, is the monopole contribution to the flow field. It is the Stokeslet flow field that results from the net force acting on the fluid.

\begin{figure}[htdp]
\includegraphics[trim = 0 150 0 50, clip=100 200 0 240, width = \linewidth]{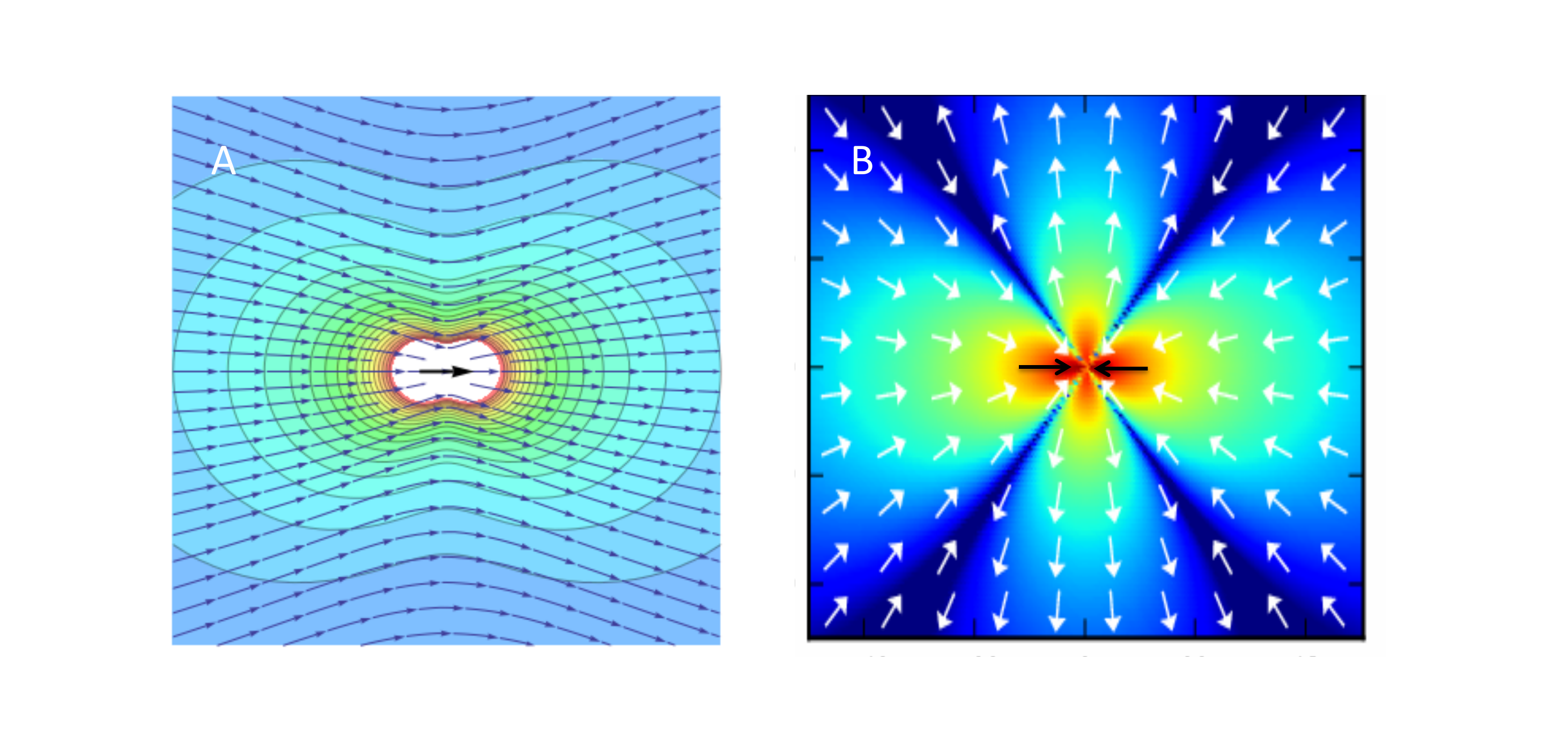}
\includegraphics[trim = 0 130 0 100, clip, width = 0.9\linewidth]{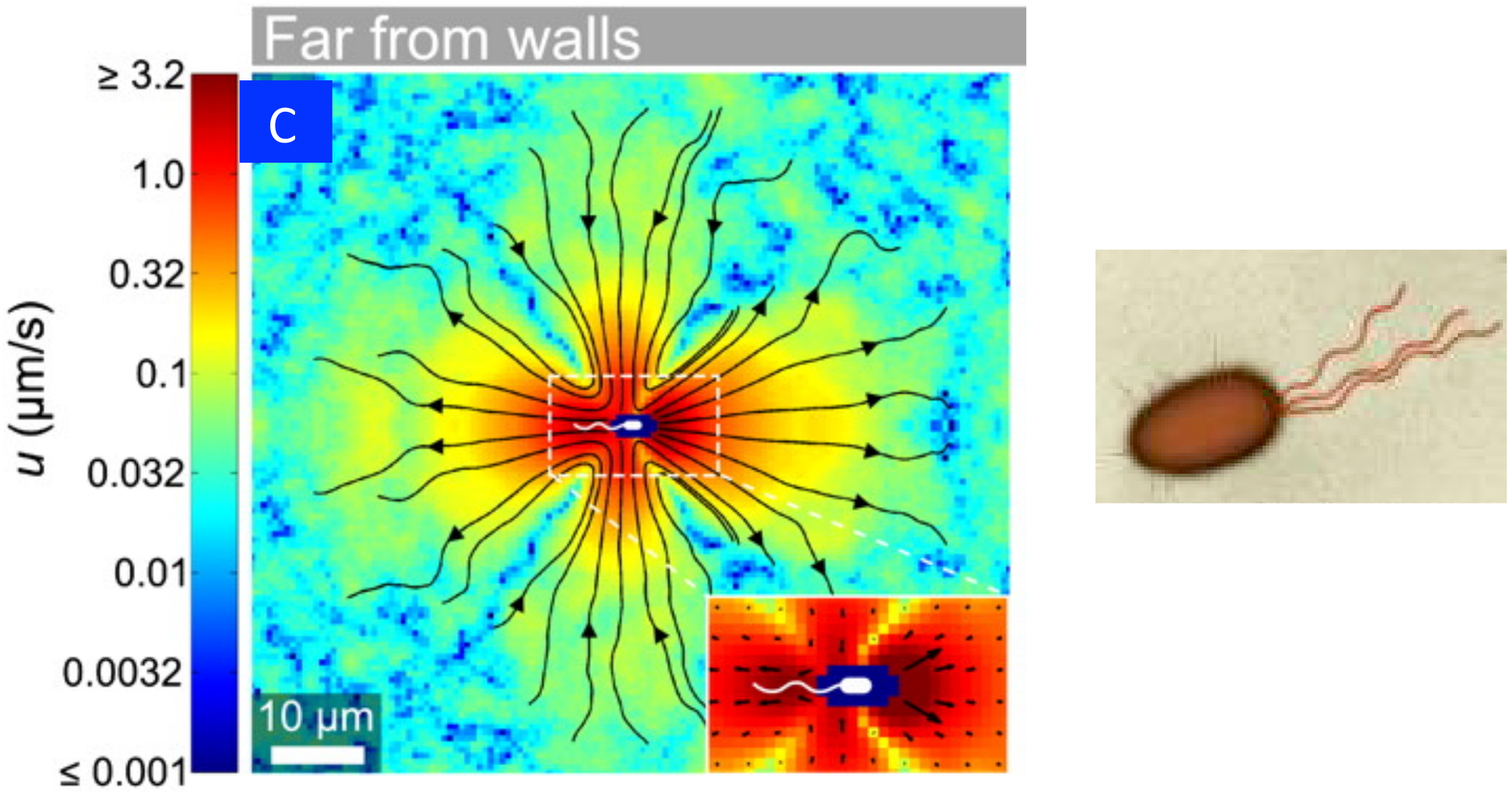}
\includegraphics[trim = 0 100 0 170, clip, width = 1\linewidth]{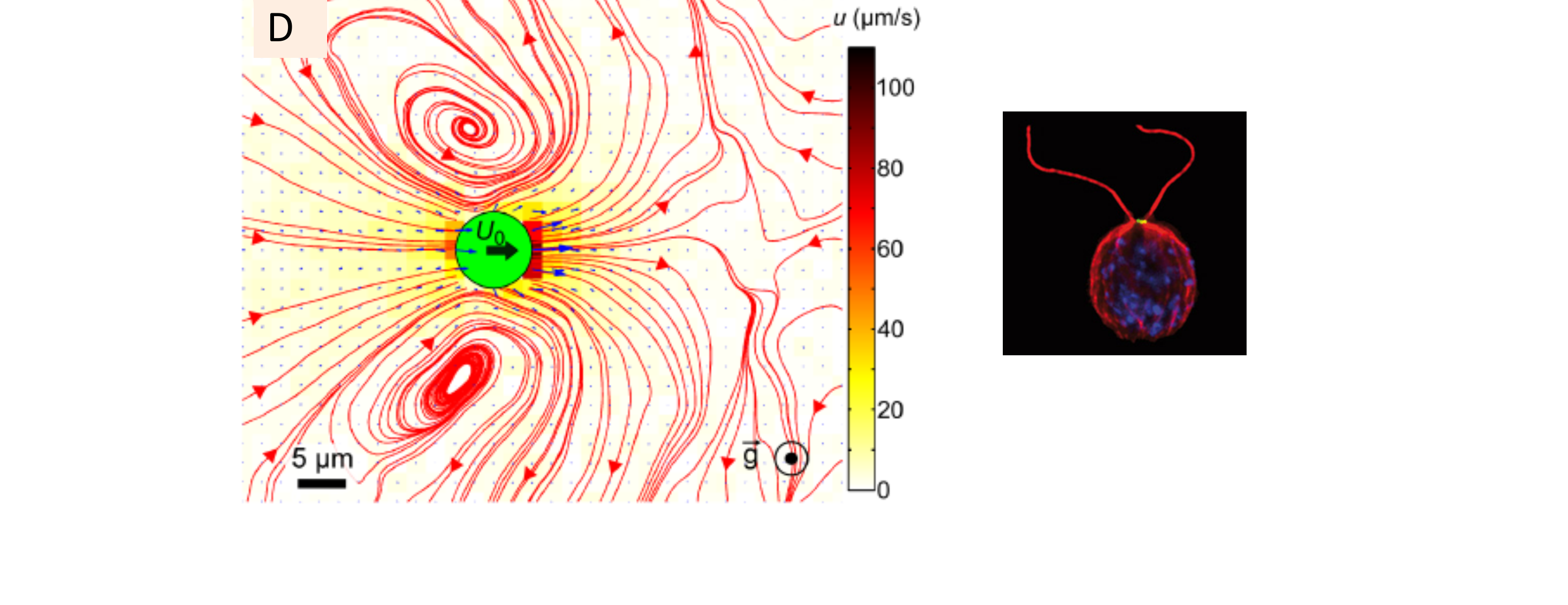}
\includegraphics[trim = 0 240 0 80, clip, width = 1\linewidth]{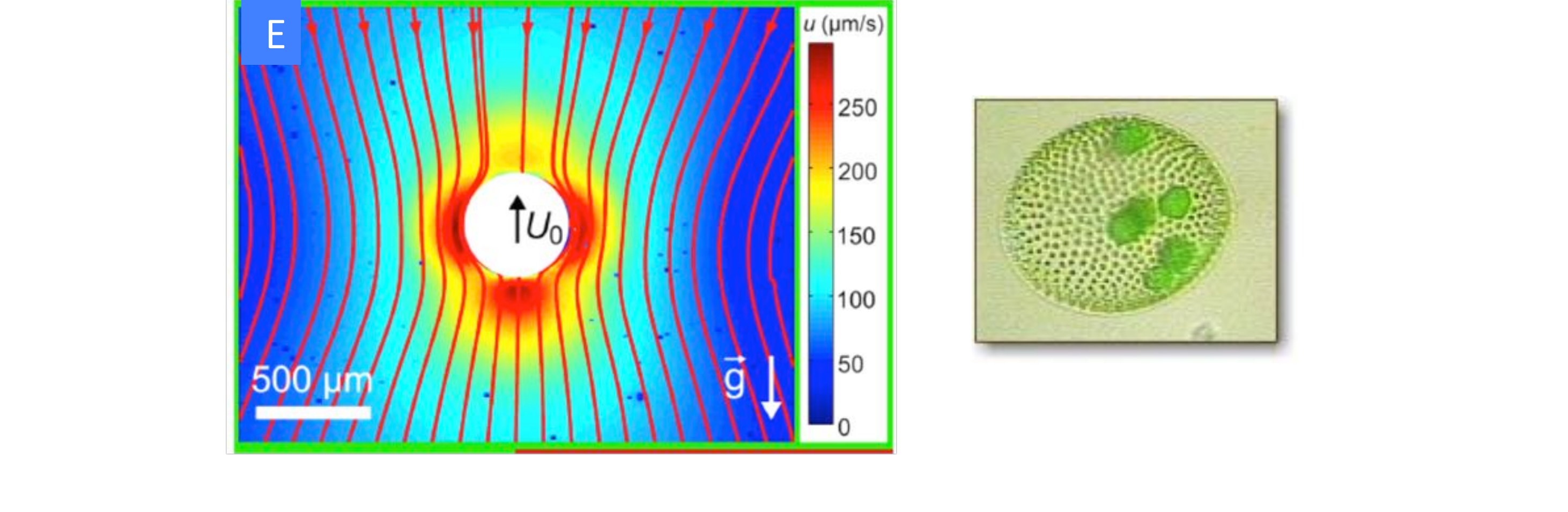}
\label{flowfig}
\caption{A: Stokeslet, the flow field of a point force. B: Stresslet, the flow field of a force dipole. Arrows indicate the flow direction, colour shading the magnitude. C: Flow field of {\it E. coli}, the far field is extensile and dipolar. D. Flow field of {\it Chlamydomonas}, the far field is contractile and dipolar. E. The flow field of {\it Volvox} is a Stokeslet (after Drescher {\it et al.}  \cite{DG10,DD11}).}
\end{figure}

We now come to a key point in the argument: A swimmer that moves autonomously can exert {\bf no net force or torque} on the fluid: all forces and torques must balance as a consequence of Newton's second law. Therefore the far flow field will, to leading order, generically be given by the next term in the multipole expansion. This is the dipolar term
\begin{equation}
 \frac{\partial G_{ij}}{\partial \xi_k} (\mathbf{r}) D_{jk},
 \end{equation}
 where
 \begin{equation}
D_{jk} = -\int \xi_k f_j \; d\mathbf{\xi}.
\end{equation}
It is conventional to write
\begin{equation}
D_{jk} -\frac{1}{3} D_{ii} \delta_{jk} \equiv S_{jk}+T_{jk},
\end{equation}
where $S_{jk}$ is a traceless symmetric tensor, referred to as the {\it stresslet}, and $T_{jk}$ is an asymmetric tensor, the {\it rotlet}. Subtracting 
$\frac{1}{3} D_{ii} \delta_{jk}$ is permissible because $\nabla \cdot \mathbf{G}=0$ and hence it can be shown that this term does not change the velocity field.

The stresslet 
\begin{equation}
S_{jk} = -\frac{1}{2} \int (\xi_k f_j  + \xi_j f_k )\; d\mathbf{\xi} + \frac{1}{3} \int \xi_i f_i \; \delta_{jk} \; d\mathbf{\xi}
\label{stresslet}
\end{equation}
is connected to straining flows and, as we shall see below, it is responsible for additional stress in  a fluid in the presence of colloids or swimmers. The rotlet term arises if the force distribution has a net torque. It is zero for microswimmers because they are torque-free.

As a concrete example, to see what a dipolar flow field looks like, consider a pair of equal and opposite co-linear forces of strength $f$ acting at $z=\pm \ell/2$. The stresslet corresponding to this force distribution is diagonal
\begin{equation}
S_{xx}=S_{yy}=  - S_{zz}/2 =\ell f/3.
\label{eq:stresslet}
\end{equation}
Some lines of algebra show that the derivative of the Stokeslet is
\begin{equation}
\frac{\partial G_{ij}}{\partial \xi_k} (\mathbf{r})=\frac{1}{8\pi\mu}\left\{ -\frac{1}{r^3} \delta_{ij} \xi_{k}
+\frac{1}{r^3}(\delta_{ik}\xi_{j} + \delta_{jk} \xi_{i})
-\frac{3}{r^5} \xi_i \xi_j \xi_k \right\}.
\label{eq:stressletflow}
\end{equation}
Using Eqs.~(\ref{eq:stresslet}) and (\ref{eq:stressletflow}) in the multipole expansion (\ref{eq:multipole}) shows that the dipolar term corresponds to a radial flow field
\begin{equation}
v_r=\frac{1}{8\pi\mu r^2}(1-3\cos^2 \theta) \ell f,
\label{dipoleflowfield}
\end{equation}
with the characteristic $r^{-2}$ dipolar dependence.

Figs.~2A and 2B compare the flow field of a Stokeslet, due to a point force, and a stresslet, due to equal and opposite forces. Note the different symmetries. For the Stokeslet there is a net velocity in the direction of the applied force whereas the dipolar field due to the stresslet has reflection symmetry about a line through the centre of the swimmer. This nematic symmetry will be important later, both in understanding the behaviour of tracer particles near a microswimmer and when I discuss the collective behaviour of active systems. Comparing Eq.~(\ref{dipoleflowfield}) to the flow field of a squirmer, Eq.~(\ref{eq:squirmer}) shows that the leading order term is a dipole of strength $(8 \pi \mu) B_2 a^2 /3$. (The constant term occurs because the velocity is written in the rest frame of the squirmer which is moving with speed $2 B_1 /3$. Higher order, quadrupolar terms, $\sim r^{-3}$, also depend on $B_1$, a pathology of this simple model.) The symmetry of the three sphere swimmer means that the dipole moment is zero for equal arm lengths, and the flow field is quadrupolar. 

In Figs.~2(C--E) we compare experimental results for three microswimmers, {\it E. coli}, {\it chlamydomonas} and {\it Volvox}  \cite{DG10,DD11}. For {\it E. coli} and {\it Chlamydomonas} the far flow field is indeed dipolar, but there are significant deviations even rather far from the swimmers due to higher order terms in the multipole expansion. For {\it E. coli} the dipolar component of the flow is directed out from the ends of the swimmer and flows towards its sides. The  {\it E. coli} is acting as a {\bf pusher} creating an {\bf extensile} flow field. For  {\it Chlamydomonas} the sign of the dipole is reversed, the swimmer is acting as a {\bf puller}, creating a {\bf contractile } flow, pumping fluid out from the sides and in to the ends of the body. The flow field of {\it Volvox}, however, has a very different symmetry. This is because it is a much bigger organism, that swims upwards against gravity. This provides a net force on the swimmer and  hence a Stokeslet term which dominates in the far field.


\section{Single swimmer hydrodynamics: applications}

Lecture 2 considers physical situations where flow fields are relevant to the behaviour of biological systems. I first discuss how a passing swimmer moves tracer particles in its vicinity, leading to simple models for the enhancement of diffusion by active particles.  I then comment on the propensity of swimmers to accumulate at surfaces and on consequences of  {\it -taxis}, the ability of some microswimmers to prefer a given direction defined by, say, gravity or a concentration gradient.

\subsection{Tracers: loops and entrainment}
Several experiments and simulations have shown that, as might be expected, bacteria enhance diffusion as a result of the flow fields they produce \cite{LG09,UH08,CL07}. I present a discussion of model systems aiming to explain how tracer particles move in a microswimmer suspension \cite{PS13,PY13,LT11}.\\
 ~\\
The instantaneous statistics of the velocity field govern the initial rate of displacement of a tracer particle. However at later times the path taken by a tracer will depend on the detailed spatial and temporal correlations of the velocity. Comparing the flow fields in Figs.~2A and B it seems likely that tracers move in a quantitatively different way as a force-free swimmer (stresslet) or as a colloid (Stokeslet) move past.

\begin{figure}[htdp]
\label{tracer}
\includegraphics[trim = 50 150 0 40, clip=50 200 0 100, width = 1.2\linewidth]{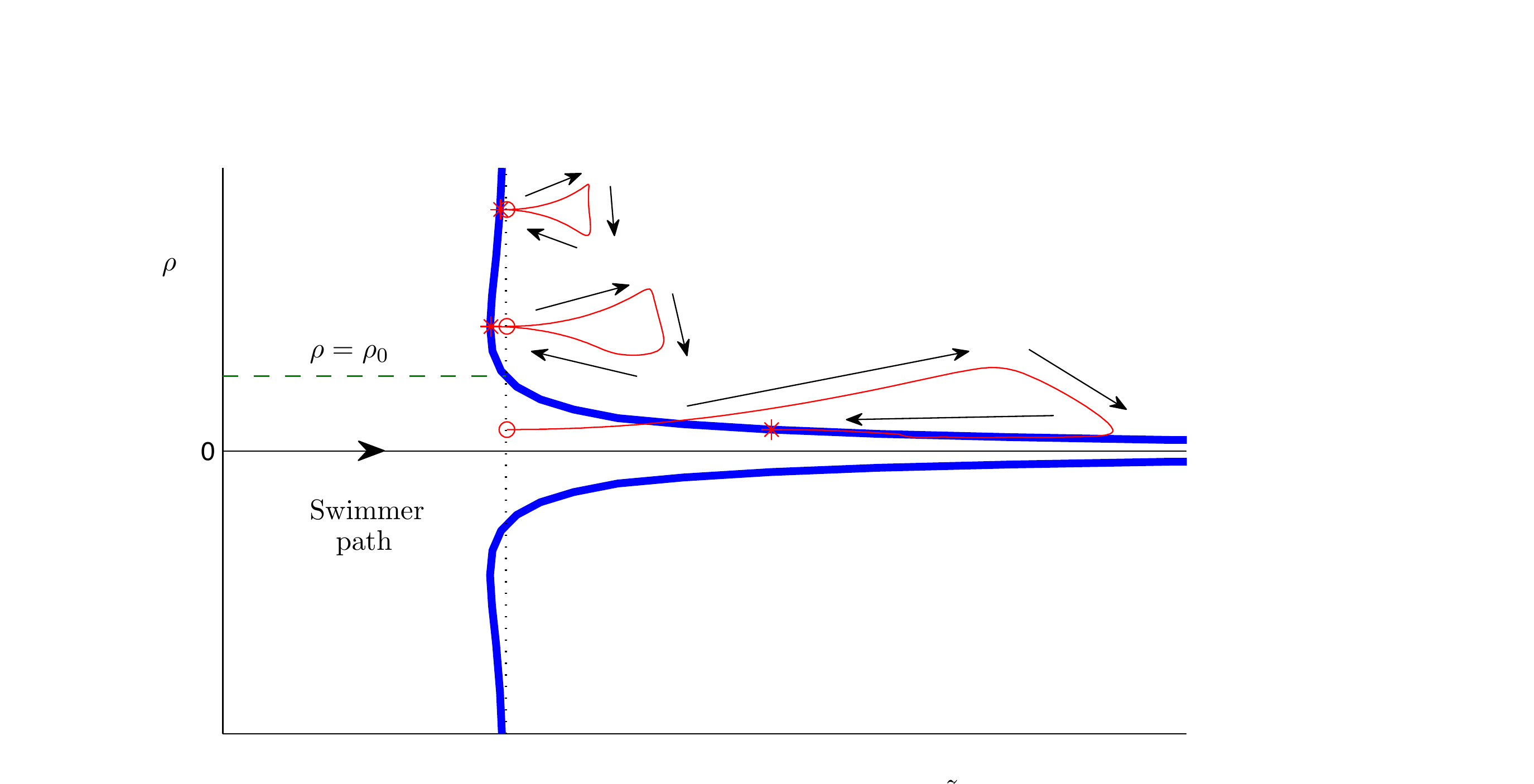}
\caption{Typical motion of a sheet of tracer particles as a swimmer moves in an infinite straight trajectory from $z = -\infty$ to $z = +\infty$ perpendicular to the sheet. The initial position of the tracer sheet is shown as a dotted line, and the envelope of the final tracer positions in blue. Examples of the loop like trajectories of the tracers are indicated as lighter, red lines (after Pushkin, Shum and Yeomans \cite{PS13}).}
\end{figure}

Consider a swimmer that moves in a straight line from $-\infty$ to $+\infty$ along the $z$-axis. The resulting displacement of a material sheet, a plane of tracers initially perpendicular to the swimmer path is shown in Fig.~3. In the figure the swimmer path is denoted by a black line, and the initial tracer positions by faint black dots. The paths of selected tracers are shown in red, and the final position of the deformed sheet of tracers in blue. These results were obtained from a numerical solution of the Stokes equations using a
boundary element method which determines the flow field and swimming velocity of the swimmer subject to no-slip boundary conditions on the swimmer surface, and the constraints that zero net force and torque act on the swimmer \cite{SG10} .

The striking feature of the tracer trajectories is their loop-like character. This is a consequence of the angular dependence of the flow field (see Fig.~2B). If the tracer is sufficiently far from the swimmer that its velocity is small compared to the swimmer velocity the loops are closed. 
This can be inferred from the symmetry of the dipolar flow field by considering how a tracer moves in the rest frame of the swimmer. The mathematical justification relies on noting that all terms except the Stokeslet in the multipole expansion are exact derivatives \cite{PS13}. 

Enhanced diffusion of the tracers in a swimmer suspension could not occur if the tracers simply moved in closed loops. However the symmetry argument neglects the Lagrangian contribution to the time derivative of the tracer velocity, which arises if the tracer moves across stream lines of the flow. 
As a result of the Lagrangian terms tracers far from the swimmer gain a net backwards displacement, whereas those close to the swimmer are displaced forwards. 
This effect becomes increasingly important for the fast-moving tracers closer to the swimmer. If a swimmer happens to pass so close to a tracer that the latter gains a velocity comparable to that of the swimmer it moves with the swimmer for a while and is said to be entrained along  the swimming direction (see Fig.~3). 

Perhaps surprisingly, the total volume of fluid displaced by the swimmer as it moves along an infinite, straight path takes a simple, universal form  \cite{D53,LP10,PS13}
\begin{equation}
v_D=\frac{4 \pi Q_\perp}{V} - v_s,
\end{equation}
where $v_s$ is the volume of the swimmer, $V$ is the swimming velocity, and $Q_\perp =- \frac{1}{2}\int f_z \rho^2 \; dS$ is a quadrupole moment with $f_z$ the $z$-component of any force on the surface of the swimmer and $\rho$ the radial co-ordinate. $v_D$ is termed the Darwin drift: by comparison, for a colloid at zero Reynolds number this quantity is infinite.

The magnitude of the Darwin drift is of the order of the swimmer's volume. Also, the characteristic entrainment length is typically of the order of the swimmer's size, $a$. These considerations allow an order of magnitude estimate of the effective tracer diffusion due to entrainment,  $D_{\mbox{entr}} $, based on a kinetic-theory approach, 
\begin{equation}
D_{\mbox{entr}}  \sim  \frac{1}{6} a^4 n V,
\label{entrainment}
\end{equation}
where we assume a dilute, uncorrelated suspension of swimmers of number density $n$.\\

Bacteria do not move along infinite straight paths because of significant rotational diffusion. Indeed some microorganisms, for example {\it E. coli}, have an explicit run and tumble behaviour which allows them to exploit a biassed random walk to move up a nutrient gradient. A finite swimmer path implies an open tracer loop and hence finite swimmer run lengths provide another important contribution to tracer diffusion. This can be estimated by using a model system introduced by Lin {\it et al.} \cite{LT11}. Their model assumes a uniform and isotropic ensemble of uncorrelated swimmers moving in straight runs of length $\lambda$, followed by instantaneous random changes of swimming direction. The consecutive runs are assumed to be statistically independent. Assuming $a \ll \lambda$ this model can be solved exactly to give a contribution to the diffusion constant from random reorientations \cite{PY13}
\begin{equation}
D_{\mbox{rr}}=\frac{4 \pi}{3} \left( \frac{\kappa}{V}\right)^2 n V
\label{rr}
\end{equation}
where $\kappa$ is a measure of the swimmer dipole strength. That this result is, surprisingly, independent of $\lambda$ is due to a fortuitous cancellation for 3D dipolar swimmers. 

In suspensions of swimmers both entrainment and random swimmer reorientations will lead to enhanced mixing and it is reasonable to assume, as a first approximation, that their effects add up. Therefore the total diffusion coefficient may be written
\begin{equation}
D=D_{\mbox{rr}}+D_{\mbox{entr}}+D_{\mbox{thermal}}
\end{equation}
where the thermal diffusion will depend on the size of the diffusing agents. Using values for {\it E. coli}: $a\sim1.4\mu$m, $\kappa/V \sim 1.45 \mu$m$^2$ in Eqs.~(\ref{entrainment}) and (\ref{rr}) gives 
$D_{\mbox{entr}}/(nV)\sim 0.6 \mu$m$^4$  and $D_{\mbox{rr}} /(nV) \sim 8.8 \mu$m$^4$ showing that random re-orientations dominate entrainment.  
These numbers are in good agreement with the experimental value  $D/(nV)=7 \pm 0.4 \mu$m$^4$ \cite{JM13}.

As an aside we note that generalisations of these arguments show that for quadrupolar swimmers in 3D and for a 2D suspension of dipolar swimmers (but with full 3D hydrodynamics) it is the contribution from entrainment that dominates the diffusion constant. I am not aware of any theoretical work on swimmers in 2D with 2D hydrodynamics but experiments on bacteria swimming in a film suggest a considerably larger diffusion constant than in 3D \cite{KG11}. 

\subsection{Swimmers in Poiseuille flow}

\begin{figure}[htdp]
\label{zottlfig}
\includegraphics[trim = 130 90 0 30, clip=50 200 0 100, width = 1.3\linewidth]{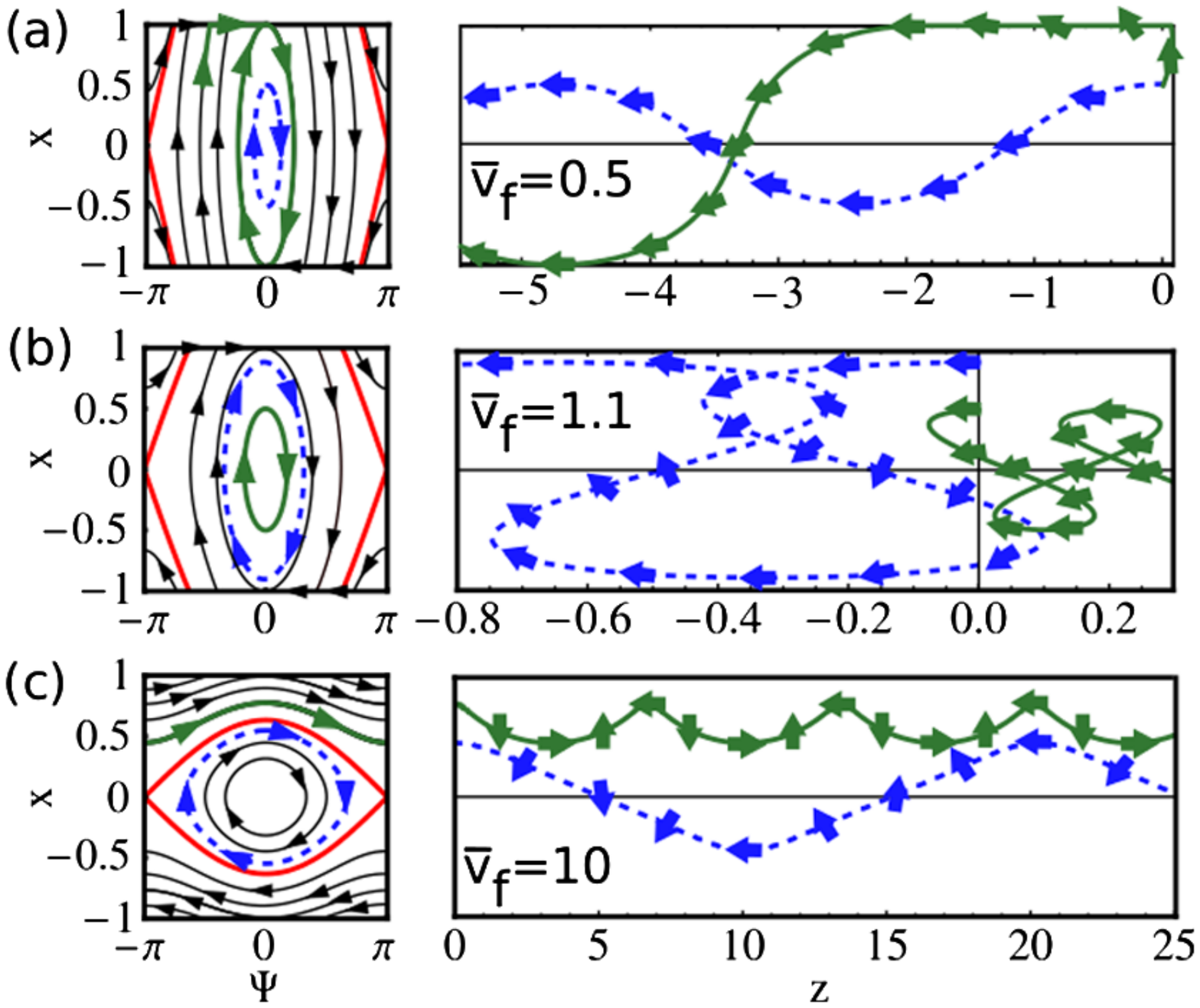}
\caption{Motion of a point swimmer in a Poiseuille flow field. The left-hand panels show the phase plots ($x$,$\psi$), with the boundary between open and oscillatory orbits indicated in red. The right hand panels show the trajectories of swimmers, starting at $z=0$ with blue and green corresponding to different initial values of $x$ and arrows indicating the instantaneous swimmer direction of motion $\hat{\mathbf{e}}$. $\bar{v}_f $ is the maximum flow speed $v_f$ scaled by the swimmer speed $v_0$. A: upstream swimming, the swimmer oscillates about the centre line. B: intermediate case, the vorticity of the flow field dominates the motion. C: downstream swimming (after Z\"ottl and Stark \cite{ZS12}).}
\end{figure}

The interplay between external flow fields and the  intrinsic swimming velocity of an active particle leads to a richness of swimming trajectories and the possibility of swimmer focussing and sorting. I discuss recent results for swimming in a Poiseuille flow, and then an effect of shear on swimmers that can sense gravity and swim upwards. \\
~\\
Z\"ottl and Stark \cite{ZS12} have shown that a simple model swimmer in Poiseuille flow can be mapped onto the equation of a non-linear pendulum. The (first) model they consider is a point microswimmer in a 2D Poiseuille flow. Even such a simple system has a surprisingly rich phase space.
Consider a swimmer at position $\mathbf{r}$ moving in direction $\hat{\mathbf{e}} $ with speed $v_0$. In an external flow $\mathbf{v}_f$ of vorticity $\mathbf{\Omega}_f=\nabla \wedge \mathbf{v}_f$ the swimmer equations of motion are
\begin{equation}
\frac{d}{dt}{\mathbf{r}}=v_0 \hat{\mathbf{e}} +\mathbf{v}_f, \quad \quad \frac{d}{dt}\hat{\mathbf{e}}=\frac{1}{2} \mathbf{\Omega}_f \wedge \hat{\mathbf{e}}.
\end{equation}
Following Zottl and Stark \cite{ZS12} we consider a 2D channel in the $x$-$z$ plane, infinite along $z$ and bounded by walls at $x=\pm 1$, and
a Poiseuille flow given by
\begin{equation}
\mathbf{v}_f=v_f(1-x^2)\hat{\mathbf{z}}
\end{equation}
where $v_f$ is the maximum flow speed in the centre of the channel. Writing the swimmer orientation in terms of a polar angle $\psi$, with the polar axis pointing upstream,
\begin{equation}
\hat{\mathbf{e}}=-\sin{\psi}\hat{\mathbf{x}}-\cos{\psi}\hat{\mathbf{z}},
\end{equation}
the equations of motion in this 2D geometry become
\begin{equation}
\dot{x}=-\sin{\psi}, \quad \quad \dot{\psi}=\frac{v_f}{v_0} x
\end{equation}
where I have rescaled time by $v_0/(\mbox{channel width})$.
Eliminating $x$ gives the pendulum equation
\begin{equation}
\ddot{\psi}+\frac{v_f}{v_0}\sin{\psi}=0.
\end{equation}
Recall that the pendulum has two solutions, an oscillation about $\psi=0$ and a circling solution. The oscillation corresponds to the swimmer moving upstream against the flow. Any deviation for the centre line is subject to a restoring torque from the vorticity and hence the swimmer trajectory oscillates around the centre of the channel. Swimming downstream any perturbation about the centre line is amplified by the vorticity, and the swimmer tumbles in the flow. For sufficiently large velocities it continues to tumble downstream, otherwise it reaches the walls and and the simple theory must be supplemented by additional physics.

As a second example consider a point microswimmer in a shear flow $v_x(z)$ such that $\mathbf{\Omega_f}=({dv_x}/{dz} ) \hat{\mathbf{y}}.$ The swimmer will tumble in the flow and trace out a cycloidal trajectory.  There is interesting behaviour if the cell also has the tendency to swim, on average, in a particular direction because it can sense eg gravity (gravitaxis), light (phototaxis) or a chemical gradient (chemotaxis) \cite{PK92}. A simple mechanism for gravitaxis occurs if the cells are bottom heavy due to an asymmetry in density or shape. 

 Putting together shear and gravitaxis (together often termed gyrotaxis) the two forces lead to an equation of motion \cite{K85}
\begin{equation}
\frac{d}{dt}\hat{\mathbf{e}}=\frac{1}{2} \{\mathbf{\Omega}_f \wedge \hat{\mathbf{e}} + \frac{1}{\beta} ( \hat{\mathbf{z}} -
(\hat{\mathbf{z}} \cdot \hat{\mathbf{e}}) \hat{\mathbf{e}} )\}
\label{gyro}
\end{equation}
where $\beta$ is a characteristic time scale that measures the time the cell takes to relax to orientation $\hat{\mathbf{z}}$. At low shear rate the steady solution to Eq.~(\ref{gyro}) is migration at an angle to $\hat{\mathbf{z}}$ given by $\sin{\theta}=\beta \Omega$. For $\beta \Omega >1$ this solution is no longer stable, and the microswimmer tumbles in the shear flow.

Durham {\it et al.} \cite{DK09} used these ideas to explain the formation of thin layers of phytoplankton in the oceans. The layers can be centimetres to metres thick and can extend horizontally for kilometres. They argue that shear gradients can be generated in the ocean by tides, wind or internal waves. As the phytoplankton swims towards the surface they reach a position where $\beta \Omega =1$ where they start to tumble and remain trapped thus forming a localised dense layer. 


\subsection{Surfaces}
I discuss why micro-organisms, such as sperm cells, {\it E. coli} and {\it Chlamydomonas} often accumulate at surfaces. 
Many experiments on micro-swimmers are performed in finite geometries, such as on microscope slides or at an interface, and bacteria often move in confined spaces in vivo or in the soil.\\
~\\
Several mechanisms have been proposed to account for the propensity of microswimmers to move near surfaces, and a combination of factors in likely to be relevant for a given organism \cite{BT08,SG10,SL12,KD13}. Before considering hydrodynamic interactions it should be appreciated that a simple self-propelled rod, and even a self-propelled sphere, will eventually hit a surface and tend to move parallel to it \cite{EW15}. Trapping on the substrate will be broken only by rotational fluctuations which allow the rod to change its swimming direction to escape from the vicinity of the surface.

Hydrodynamic interactions with the wall can be taken into account by considering an image swimmer at a position corresponding to the reflection of the swimmer in the wall. For free boundaries this results in adding a term to each Stokeslet corresponding to a Stokeslet of equal magnitude but pointing in the opposite direction
\begin{equation}
\label{blakefree}
\mathcal{B}_{ij}(\mathbf{r^*}) =- \frac{1}{8\pi\mu} 
 \left( \frac{\delta_{ij}}{r^*} + \frac{r^*_i r^*_j}{r^{*3}}\right),
\end{equation}
where $\mathbf{r}^{*}$ is the distance from the image to the point at which the flow is calculated. For a no-slip boundary a more complicated image system is required to preserve the no-slip condition. The necessary additional terms in the Greens function, often called the Blake tensor, are \cite{B71a}
\begin{eqnarray}
\label{blakenoslip}
\mathcal{B}_{ij}(\mathbf{x},\mathbf{r^*}) = \frac{1}{8\pi\mu} 
\left\{
- \left( \frac{\delta_{ij}}{r^*} + \frac{r^*_i r^*_j}{r^{*3}}\right)
+ 2 H \mathbf{M}_{jl}
\frac{\partial}{\partial r^*_l} 
\left( 
\frac{H r^*_i}{r^{*3}} - 
\left( \frac{\delta_{ix}}{r^*} + \frac{r^*_i r^*_x}{r^{*3}} \right)
\right)
\right\},
\end{eqnarray}
where $H$ is the distance of the swimmer from the wall, and the mirror matrix is $\mathbf{M} = \mbox{diag}(-1,1,1)$ for the boundary along $x$. For a Stokeslet oriented parallel (perpendicular) to the boundary the Blake image system comprises three parts. The first is again a Stokeslet of equal magnitude but pointing in the opposite direction to the physical Stokeslet. The second is quadrupolar, a source doublet oriented in the opposite (same) direction with magnitude $2H^2$. The third contribution  is an asymmetric dipole, sometimes called the Stokes doublet  with magnitude $2H$. 

However hydrodynamic interactions are not the only contribution and, following early work by Berke {\it et al.} \cite{BT08},  a lot of effort has been devoted to trying to unravel the way in which swimmers interact with surfaces, e.g., Refs.\cite{SG10,SL12,KD13}. 
The conclusion  is that details of the swimmer play an important role in determining its behaviour. Whether or not a microswimmer tends to move towards or away from a wall depends on its aspect ratio and on the details of its activity. Those swimmers that tend to move towards the surface usually have sufficiently small rotational diffusion that they will eventually hit the wall and both high order terms in the multipole expansion and steric interactions will become important. 

For microswimmers such as {\it E. coli} the cell head and flagella spin in different directions so that the system is torque free in an infinite fluid. However, individual cells are often observed to swim in circles close to boundaries.  This is because there is additional drag on the side of the bacterium facing the wall if the bacterium approaches a wall leading to a torque that moves the swimmer in a clockwise orbit for a no-slip boundary and an anti-clockwise orbit for a slip boundary. More details of the behaviour of microswimmers near boundaries are clearly explained and references are listed in the recent review by Elgeti {et al.} \cite{EW15}.


\section{Collective hydrodynamics of active entities}

In lecture 3 I will write down and motivate the continuum equations of motion that describe the hydrodynamics of dense active nematics, such as suspensions of microtubules driven by molecular motors and dense collections of microswimmers. The equations are based on those of passive liquid crystals with an additional term in the stress resulting from the activity. This term leads to a hydrodynamic instability which drives the nematic state unstable and leads to active turbulence.

\subsection{Nematic liquid crystals}

\begin{figure}[htdp]
\label{lc}
\includegraphics[trim = 80 20 0 30, clip=100 200 0 100, width = 1.1\linewidth]{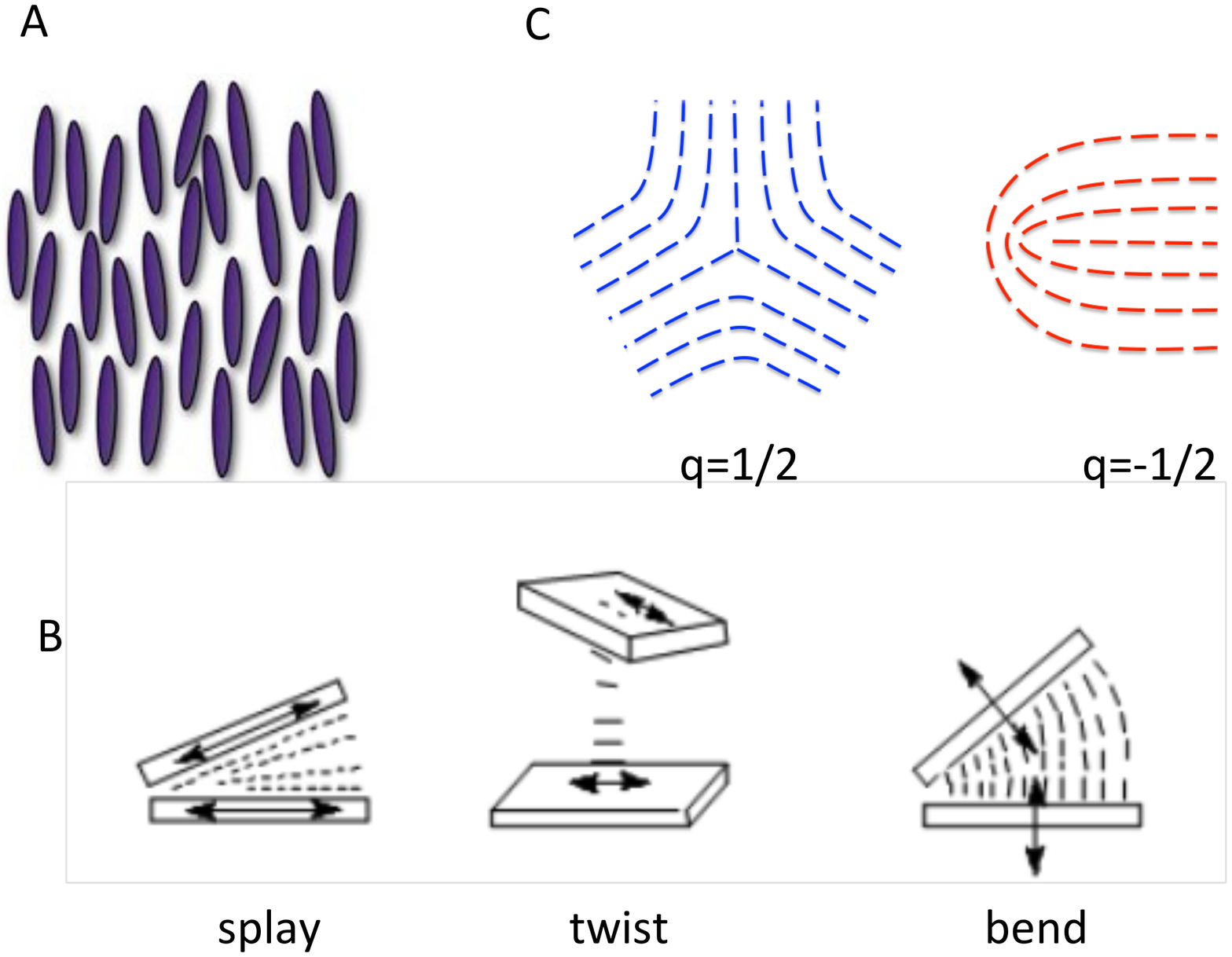}
\caption{A: Particles in the nematic state have long-range orientational order but no long-range positional order. B: A 3D liquid crystal has three elastic constants which quantify the energy associated with splay, twist and bend distortions (after de Gennes and Prost \cite{GP95}). C: Topological defects in 2D.}
\end{figure}

We introduce liquid crystals and summarise properties that are relevant to the description of active systems. After describing the nematic state we define an associated tensor order parameter $\bf{Q}$. The elastic properties of the nematic are considered, and we introduce the concept of topological defects. \\
~\\
Liquid crystals are fluids where the constituent molecules are anisotropic, taking usually rod-like, but sometimes disc-like, shapes \cite{GP95,C92}. At high temperatures there is no long-range order, but as the temperature is lowered there is a weakly first order transition to a nematic state. This is characterised by orientational long-range order with the molecules lining up, on average, in a given direction but with no long-range order in the molecular positions (Fig.~5A).  The transition is driven by entropy: alignment means that the molecules have more space to move around.

The direction of each molecule, called the director $\mathbf{n}$, can be coarse grained to give an order parameter describing the mean molecular orientation. Note that the director is a `headless' vector, with no distinction between its ends. A fuller description of the nematic state which incorporates the magnitude of the orientational order (and allows for the possibility of describing secondary ordering perpendicular to the director) is the 3-dimensional traceless tensor $\bf{Q}$ defined by
\begin{equation}
Q_{ij} = \frac{3q}{2} \langle   n_{i}n_{j}-\delta_{ij}/3 \rangle.
\end{equation}
For example, for ordering of magnitude $q$ along the $z$-axis,  $Q_{\alpha\beta}$ is a diagonal matrix with entries $(-q/2,-q/2,q)$. The Landau-de Gennes expansion for the bulk free energy density of a nematic liquid crystal is
\begin{equation}
\mathcal{F}_{\mbox{bulk}} =  A Q_{ij} Q_{ji}/2 + B Q_{ij} Q_{jk} Q_{ki}/3 + C (Q_{ij} Q_{ji})^2/4
\label{eq:free}
\end{equation}
where $A$, $B$, $C$ are material parameters. This predicts a first order transition to the nematic state at $AC/B^2=1/27$.

In the nematic state it is favourable for the directors to align parallel and any deviations from this ordering results in an additional, elastic contribution to the free energy. There are three distinct types of bulk distortion in three dimensions, splay, twist and bend, illustrated in Fig.~5B. These are associated with elastic constants $K_1$, $K_2$ and $K_3$ respectively and can be written as gradient terms in the Landau expansion of the free energy, most transparently in terms of the director field $\mathbf{n}$
\begin{equation}
F_{\mbox{el}}({\mathbf{n}})=\frac{K_1}{2} (\nabla \cdot \mathbf{n})^2 + \frac{K_2}{2} (\nabla \cdot \mbox{curl} \;\mathbf{n})^2
 + \frac{K_3}{2} (\nabla \wedge \mbox{curl} \;\mathbf{n})^2
\label{eq:elasticfreen}
\end{equation}
or, in terms of $Q$,
\begin{equation}
F_{\mbox{el}}({\mathbf{Q}})= \frac{L_1}{2} (\partial_k Q_{ij} )^2 + \frac{L_2}{2} (\partial_j Q_{ij} )(\partial_k Q_{ik} ) +\frac{L_3}{2} Q_{ij}
(\partial_iQ_{kl} )(\partial_j Q_{kl}).
\label{eq:elasticfreeQ}
\end{equation}
In two dimensions there are only bend and splay distortions. 

As an aside, Eqs.~(\ref{eq:elasticfreen}) and (\ref{eq:elasticfreeQ}) illustrate the arbitrary nature of the Landau expansion. Physically there are three long wavelength elastic distortions and there are three terms in gradients of $\mathbf{n}$ squared in the expansion in $\mathbf{n}$. However, there are only two terms in gradients of $\bf{Q}$ squared in the expansion in $\bf{Q}$ and a term of higher order in $\bf{Q}$
must be included to allow all three elastic constants to be varied independently. 

A second consequence of the symmetry of the nematic state is the existence of topological defects, singular distortions of the director field that cannot relax to equilibrium without rearrangement of the molecules at infinity. Topological defects in two dimensions are shown in Fig.~5. 
A single defect has an infinite energy in an infinite system, so can only exist due to a boundary or (real) defect. Therefore defects are usually formed in pairs, typically following a quench from the isotropic to the nematic phase. In the absence of pinning sites defects with opposite changes attract, approach each other and annihilate. 


\subsection{Beris-Edwards equations}

We describe the continuum equations of motion of a nematic liquid crystal, written in terms of the $\bf{Q}$-tensor. These are more complicated than the Navier-Stokes equations for a simple fluid because of the shape of the molecules: they both respond to flow gradients, and their motion induces backflows.\\
~\\
The Beris-Edwards equations of nematohydrodynamics are \cite{B94}
\begin{align}
\partial_t \rho + \partial_i (\rho u_i)&=0,
\label{eqn:cont}\\
\rho (\partial_t + u_k \partial_k) u_i &= \partial_j \Pi_{ij},
\label{eqn:ns} \\
(\partial_t + u_k \partial_k) Q_{ij} - S_{ij} &= \Gamma H_{ij}.
\label{eqn:lc}
\end{align}
Eq.~(\ref{eqn:cont}) is the continuity equation, Eq.~(\ref{eqn:ns}), the generalisation of the Navier-Stokes equation to the liquid crystalline system, describes the evolution of the velocity field and  Eq.~(\ref{eqn:lc}) is a convection-diffusion equation for the dynamics of the order parameter field. The notation we use here allows us to write the equations in a compact way, but each term encapsulates important details of the hydrodynamics of this viscoelastic fluid.

Consider first the evolution of the order parameter field,  Eq.~(\ref{eqn:lc}). As a result of their anisotropy, liquid crystal molecules rotate in velocity gradients. This is accounted for by the generalised advection term
\begin{equation}
S_{ij} = (\lambda E_{ik} + \Omega_{ik})(Q_{kj} + \frac{\delta_{kj}}{3}) + (Q_{ik} + \frac{\delta_{ik}}{3})
 (\lambda E_{kj} - \Omega_{kj}) - 2 \lambda (Q_{ij} + \frac{\delta_{ij}}{3})(Q_{kl}\partial_k u_l)
 \label{eqn:cor}
\end{equation}
where
\begin{equation}
E_{ij} = (\partial_i u_j + \partial_j u_i)/2, \quad \quad \quad \quad \Omega_{ij} = (\partial_j u_i - \partial_i u_j)/2
\end{equation}
are the strain rate and vorticity tensors.  
$\lambda$ is the alignment parameter which determines whether the collective response of the nematogens to a velocity gradient is dominated by the strain or the vorticity. The director aligns at a given angle to a shear flow, called the Leslie angle, if
\begin{equation}
\mid \lambda \mid\;\; > \frac{9q}{3q+4}.
\end{equation}
If this equality is not satisfied the molecules rotate (tumble) in the flow. $\lambda$ depends on the shape of the particles with $\lambda>0$ and $\lambda<0$ corresponding to rod-like and plate-like particles respectively.  

The molecular field
\begin{equation}
H_{ij} = -\frac{\delta \mathcal{F}}{\delta Q_{ij}} + \left(\frac{\delta_{ij}}{3}\right) {\rm Tr} \left(\frac{\delta \mathcal{F}}{\delta Q_{kl}}\right)\label{eqn:molpot}\\
\end{equation}
ensures that the system relaxes diffusively to the minimum of a free energy, e.g., $\mathcal{F}_{\mbox{bulk}}+\mathcal{F}_{\mbox{el}}$ from Eqs.~(\ref{eq:free}) and (\ref{eq:elasticfreeQ}). The diffusion constant $\Gamma$ controls the time scale over which the relaxation occurs.

Turning now to the momentum equation~(\ref{eqn:ns}), the inertial terms on the left-hand side can be neglected at low Re. On the right-hand side $\Pi_{ij}$ is the stress tensor. It is a sum of the usual viscous stress
\begin{equation}
\Pi_{ij}^{viscous} = 2 \eta E_{ij}
\end{equation}
and an elastic stress
\begin{eqnarray}
\Pi_{ij}^{passive}&=&-P\delta_{ij} + 2 \lambda(Q_{ij} + \frac{\delta_{ij}}{3}) (Q_{kl} H_{lk})
-\lambda H_{ik} (Q_{kj} + \frac{\delta_{kj}}{3}) \nonumber\\ 
& &-\lambda (Q_{ik} + \frac{\delta_{ik}}{3}) H_{kj}
-\partial_i Q_{kl} \frac{\delta \mathcal{F}}{\delta \partial_j Q_{lk}} 
+ Q_{ik}H_{kj} - H_{ik} Q_{kj}.
\label{elasticstress}
\end{eqnarray}
The elastic stress is a second consequence of the anisotropic nature of the liquid crystal molecules. It occurs because, as the molecules turn they induce a shear flow, often called the `back-flow'. In general back-flows are small, and make a qualitative rather than quantitative difference to the hydrodynamic fields of liquid crystals.


\subsection{Adding activity}
We write down the continuum equations of motion for an active nematic and motive the form of the active term. We show that vortical flows destabilise the nematic order whereas straining flows can help to increase its magnitude.\\
~\\ 
A simple extension to the Beris-Edwards equations suffices to describe an active nematic \cite{HR04}. This is an additional term in the stress tensor
\begin{equation}
\Pi_{ij}^{active} = -\zeta Q_{ij}
\label{activestress}
\end{equation}
proportional to the $\bf{Q}$ tensor and a coefficient $\zeta$ that measures the strength of the activity. $\zeta>0$ describes an extensile system and $\zeta<0$ a contractile system. 

The form of the active term can be motivated by realising that the contribution to the continuum stress tensor from the 
dipole terms in the multipole expansion is the averaged stresslet per unit volume \cite{B70} which, from Eq.~(\ref{stresslet}), is
\begin{equation}
\Pi_{jk} = \langle-\frac{1}{2} (\xi_k f_j  + \xi_j f_k )+ \frac{1}{3} \xi_i f_i \delta_{jk} \rangle.
\end{equation}
Assuming that each of the active particles produces forces along the director $f_j= \pm f n_j$ at positions $\xi_k=\pm a n_k$ this becomes
\begin{equation}
\Pi_{jk}=\langle fa (-\frac{1}{2} (n_k n_j  + n_j n_k )+ \frac{1}{3} n_i n_i \delta_{jk}) \rangle =\langle fa (-n_j n_k +\frac{1}{3}\delta_{jk}) \rangle \equiv -\zeta {Q}_{jk}
\label{actmapping}
\end{equation}
averaging over a region where the dipole moment $fa$ and the magnitude of the nematic order $q$ can be considered constant. The relation~(\ref{actmapping})  makes it apparent that $\zeta$ is a measure of the activity, and that the sign of the stress distinguishes extensile (positive) and  contractile (negative) particles.

\begin{figure}[htdp]
\label{unstableinst}
\includegraphics[trim = 50 80 0 50, clip=50 200 0 100, width = 1.1\linewidth]{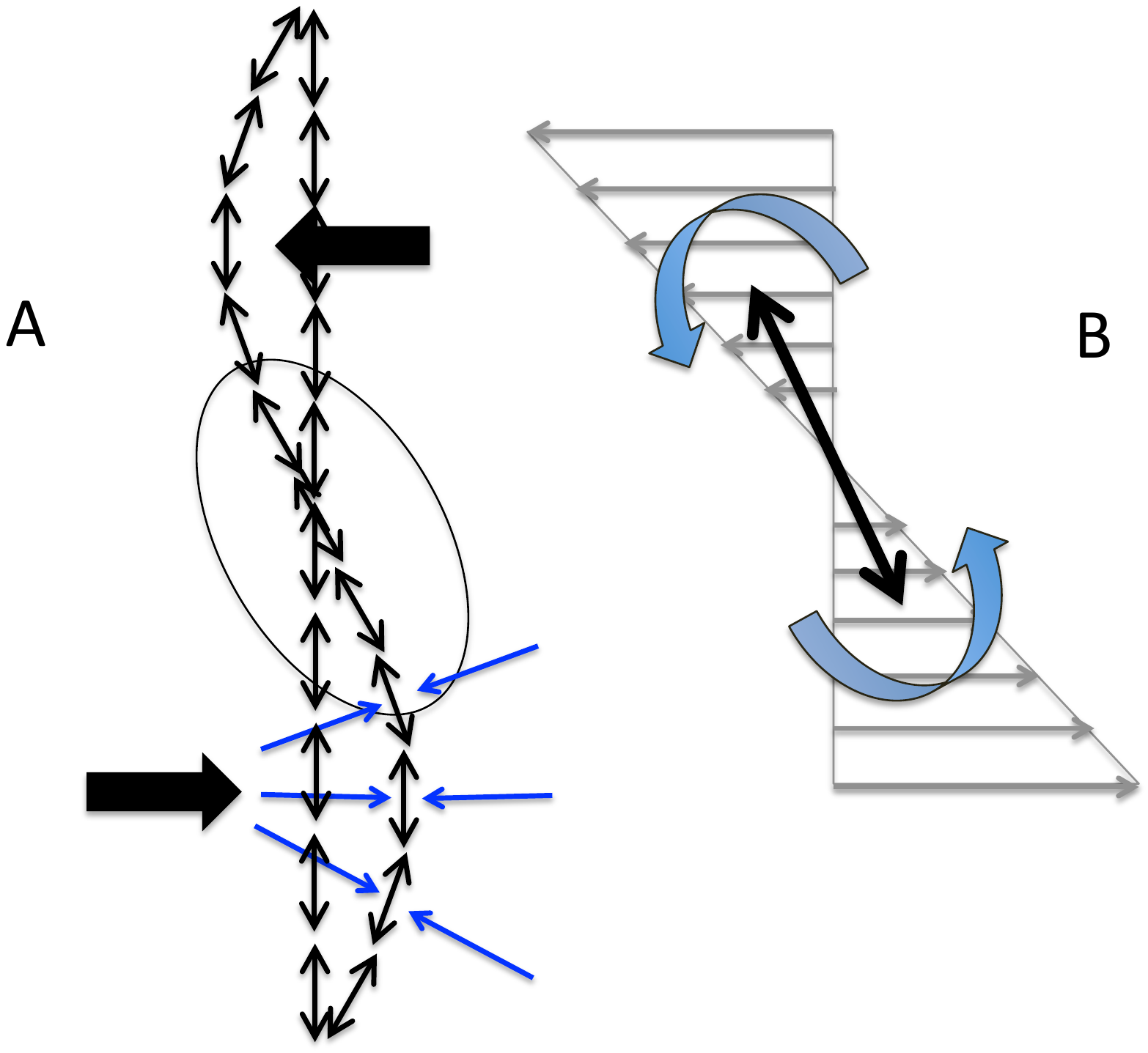}
\caption{An extensile active nematic is unstable to a bend instability. A: each nematogen in the bent configuration produces flow shown in blue. The flow is unbalanced giving a resultant vortical flow shown as large black arrows. B: The shear acts to increase the bend deformation destabilising the ordered nematic state (courtesy of S. Thampi).}
\end{figure}

Many properties of active nematics can be interpreted by noting that the active term appears in the stress, under a derivative, and therefore {\bf any change in the direction or orientation of the nematic order induces a flow}. An immediate consequence is that the active nematic state is unstable to vortical flow \cite{SR02,VJ05}. The physics behind the instability is illustrated in Fig.~6. For an extensile system a bend perturbation of the nematic director leads to a flow field indicated by the blue arrows. These are closer together on the concave side of the bend, a pictorial representation that the flow is stronger here, and thus the resultant flow is as shown by the large black arrows. This is a rotational flow that tends to turn the director further from its equilibrium position (Fig.~6B) thus destabilising the nematic state. In a similar way contractile systems are unstable to splay instabilities \cite{EY09}.

\begin{figure}[htdp]
\label{stableinst}
\includegraphics[trim = 40 250 0 30, clip=50 200 0 200, width = 1\linewidth]{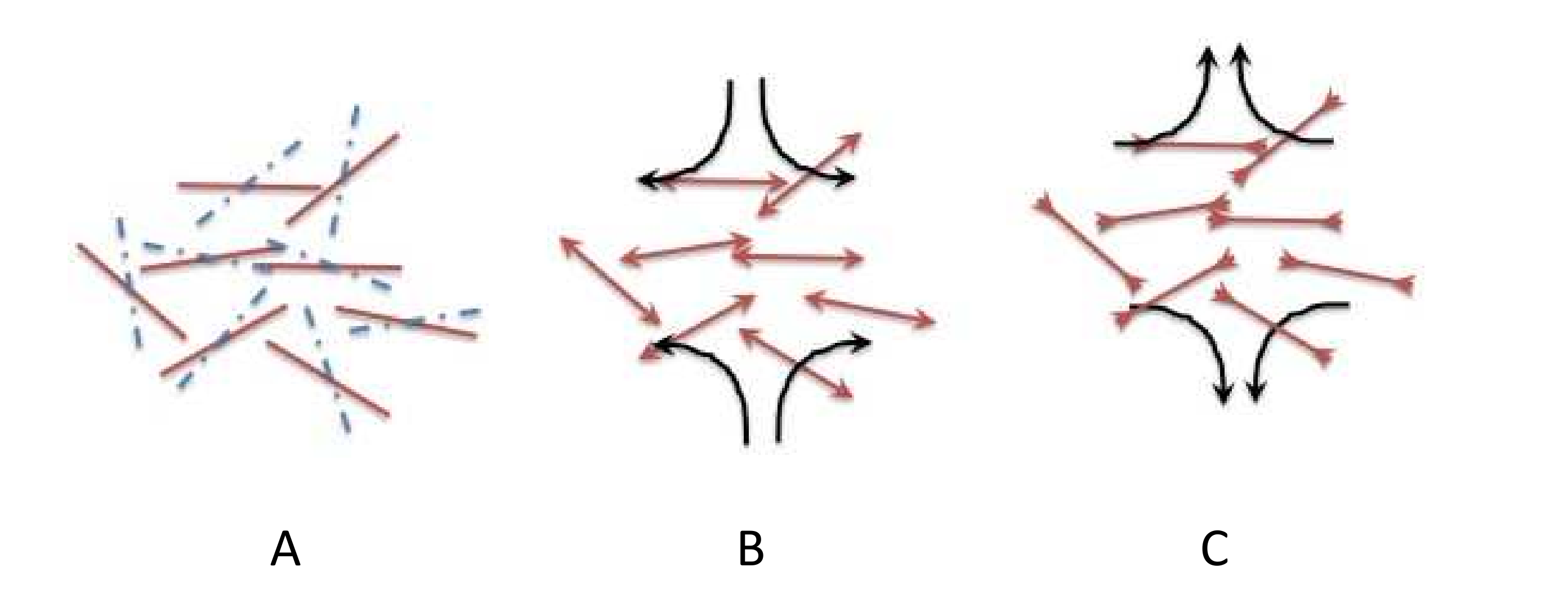}
\caption{Activity can act to stabilise nematic ordering in a suspension of rods. A: A fluctuation that results in local nematic order. B: There is a resulting extensile flow field (black) that enhances the nematic order. C: In a contractile system the flow field acts to destroy the order (after Thampi {\it et al.} \cite{TD15}). }
\end{figure}

By contrast extensional flow can tend to stabilise nematic ordering
(Fig.~7) \cite{TD15}. Recall that a rod ($\lambda>0$) attains a stable position in an extensional flow when it is aligned along the extensional axis. Starting from an isotropic configuration of elongated particles consider a fluctuation that induces a small ordering, along a given direction as in Fig.~7A. If the particles are active they generate dipolar flow fields, which enhance the instantaneous order for an extensile system  or destabilise it in the contractile case as is apparent from  Figs.~7B and C respectively. It can be shown that the response of the $\mathbf{Q}$-tensor to the extensional part of the flow generates an effective molecular potential and, in extensile systems, the hydrodynamics acts as an effective force favouring the alignment of the active rods. 
For plate-like particles, $\lambda<0$, however, the stable position is along the compressional axis with plate orientation defined normal to the plate. Therefore now it is contractile stress that favours ordering.  Since these mechanisms depends only on the extensional part of the flow field, they hold for both aligning and tumbling particles.


\section{Collective hydrodynamics: applications}

We have seen that the active nematic state is unstable to flow and in Lecture 4 we describe {\bf active turbulence}, the non-equilibrium steady state that replaces the nematic order.  We first use simulations to motivate our description of the mechanisms driving active turbulence and then describe an experimental system, a mixture of microtubules and molecular motors, where active turbulence is observed. We discuss active anchoring at an active--passive interface.

\subsection{Active turbulence}

We describe how an interplay between defects and walls in the director field and jets and vortical structures in the velocity field leads to a turbulent-like state in active nematics.\\
~\\
\begin{figure}[htdp]
\label{activeturbexamples}
\includegraphics[trim = 60 0 0 30, clip=50 200 0 200, width = 1.2\linewidth]{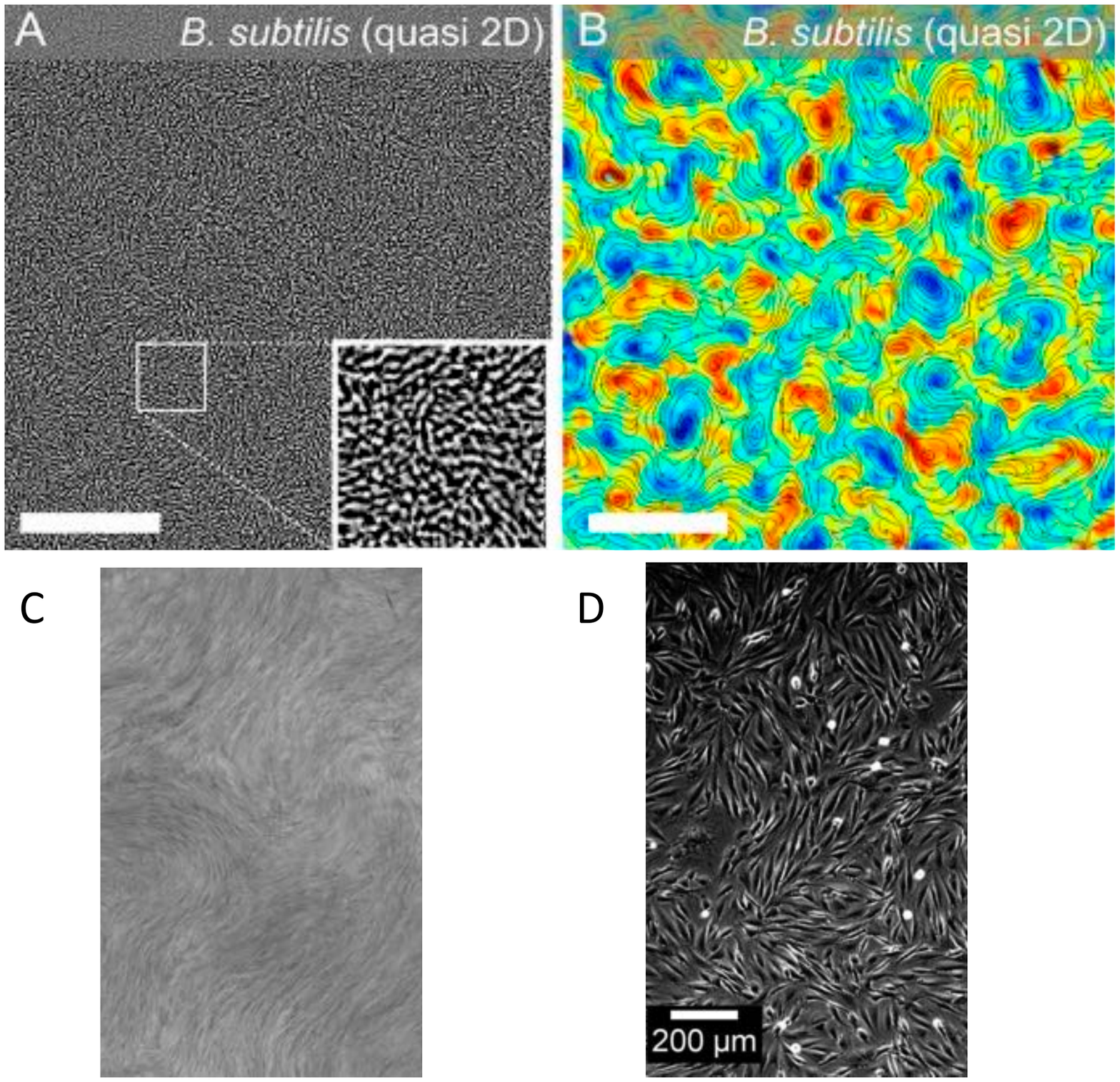}
\caption{Examples of active turbulence. A: swimming bacteria. B: the corresponding vorticity field. Red (blue) colouring corresponds to regions of high positive (negative) vorticity. Vortices are typically $\sim 5$ bacteria lengths in diameter (after Wensink {\it et al.} \cite{WD12}). C: swimming bacteria in a liquid crystal (after Zhou {\it et al.} \cite{ZS14}). D: fibroblast cells (courtesy of G. Duclos).  }
\end{figure}
Active turbulence occurs in dense active nematics \cite{DC04,WD12,SS12,MJ13,TG13,TG14,TG14a,G14}. It is a state characterised by a flow field  with continually changing regions of high vorticity. Hence the name -- active turbulence or mesoscale turbulence -- that comes about because of a superficial visual resemblance to inertial turbulence. However the latter is a consequence of high Reynolds number; here we are at low Reynolds number and the physics is very different. Active turbulence has been observed in a wide range of systems, at different length scales, examples of which are shown in Fig.~8. 

\begin{figure}[htdp]
\label{actturb}
\includegraphics[trim = 20 80 0 20, clip=50 200 0 200, width = 0.9\linewidth]{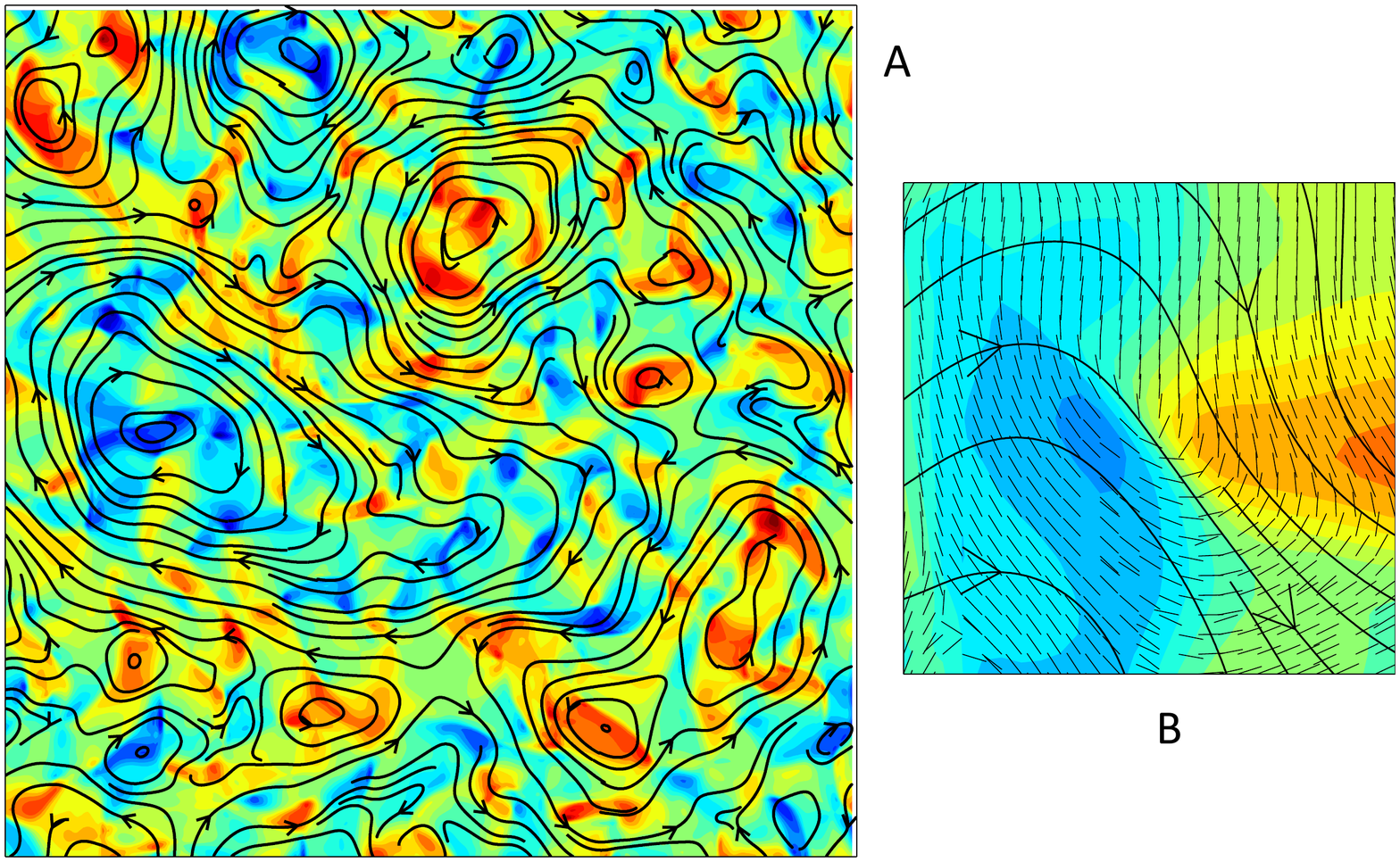}
\includegraphics[trim = 20 0 0 60, clip=50 200 0 200, width = 0.9\linewidth]{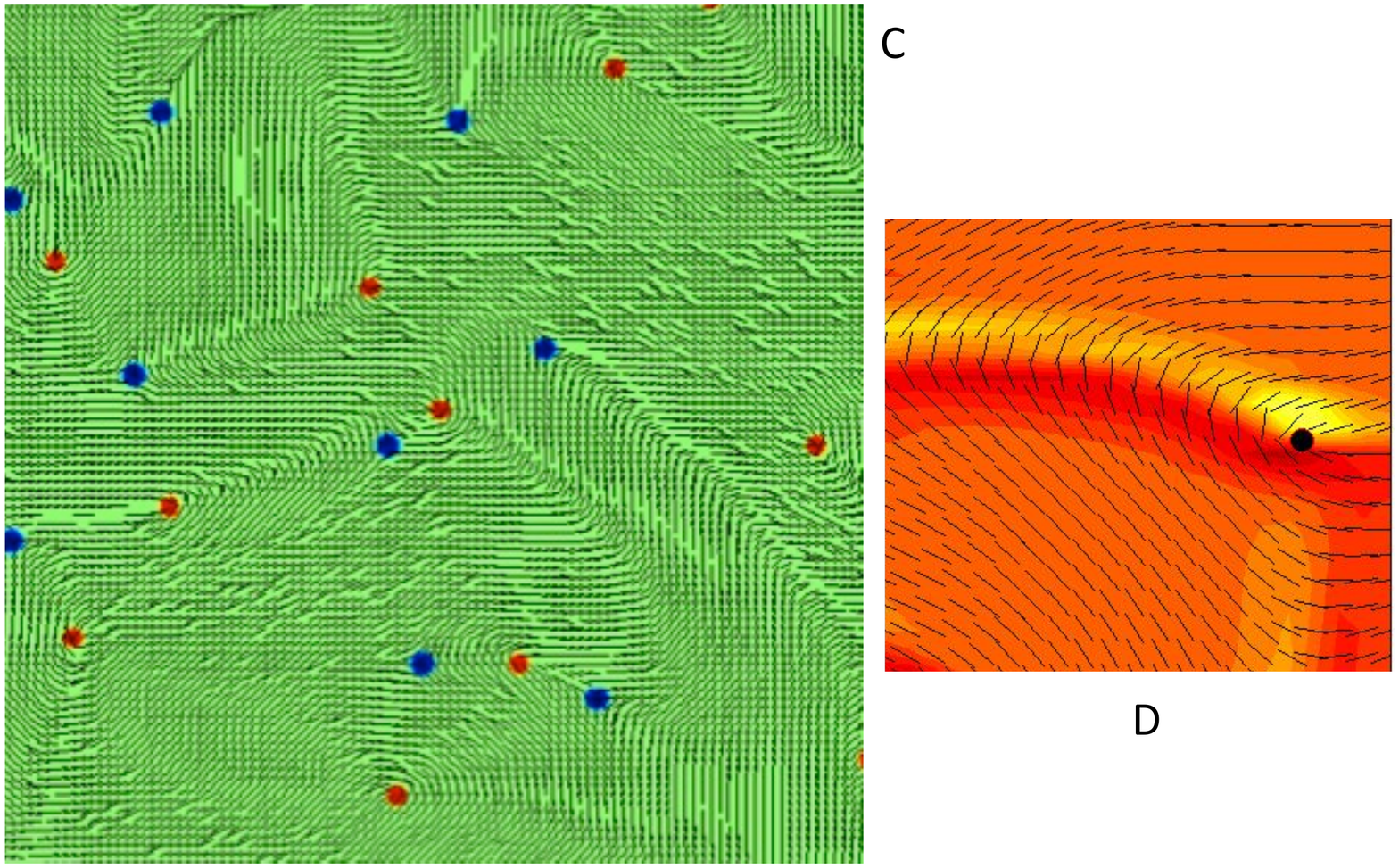}
\includegraphics[trim = 20 50 0 220, clip=50 200 0 260, width = 0.9\linewidth]{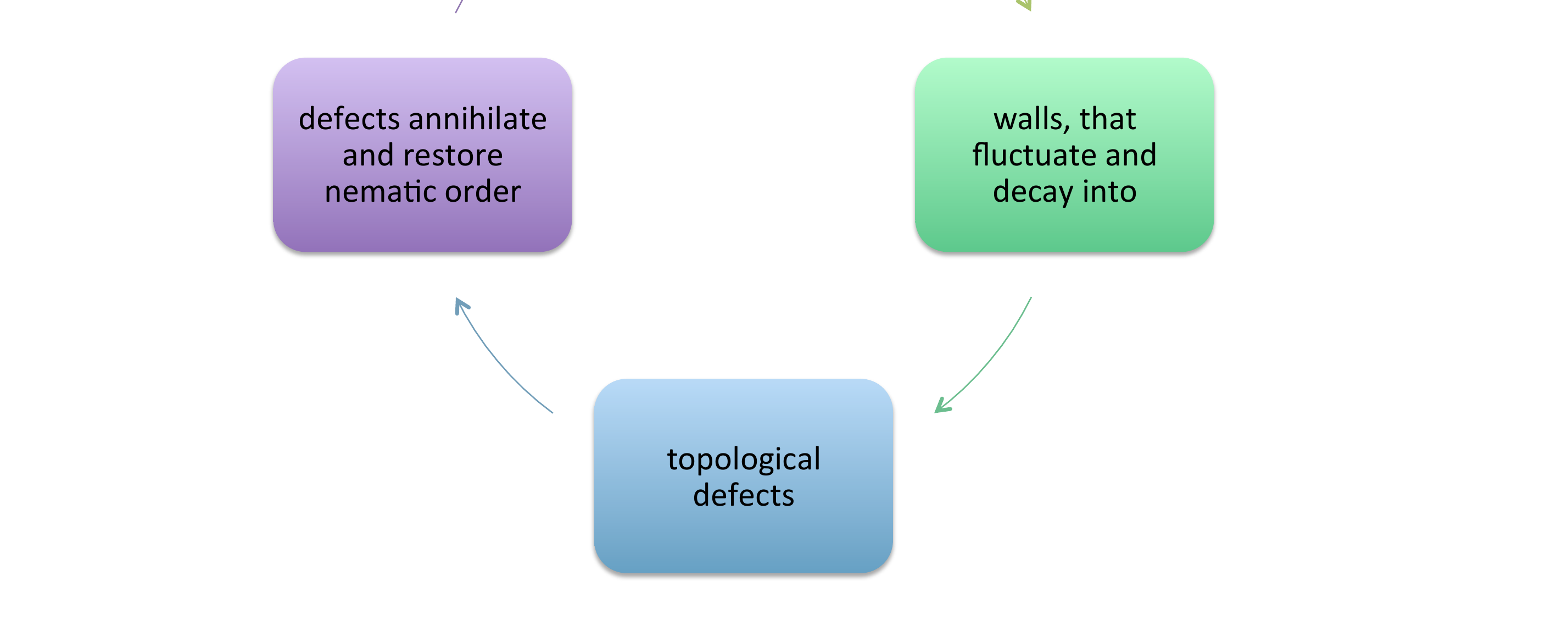}
\caption{Active turbulence. A: Velocity field showing the flow lines and a colour plot of the vorticity. Red indicates strong positive vorticity and blue strong negative vorticity. B: Close up of a region of A showing the director field to illustrate how velocity jets are driven by bend instabilities. C: Director field showing $\pm 1/2$ topological defects in red and blue. The lighter lines are walls, linear regions of high bend. D: Close up of a wall (after Thampi {\it et al.} \cite{TG13,TG14}). E: Schematic representation of the mechanism leading to active turbulence.}
\end{figure}

Fig.~9A, obtained by numerically integrating Eqs.~(\ref{eqn:cont})--(\ref{eqn:lc}),  shows an instantaneous snapshot of the flow field in the active turbulent state. Black lines are streamlines of the flow, and red and blue indicate regions of high and low vorticity. This is a dynamic pattern which 
changes in time, but is a statistical steady state. Fig.~9B magnifies a small region of this plot to show the director field. Note the bend distortion of the director which induces a stress, and hence a jet-like flow, which creates areas of positive and negative vorticity on either side of the jet. This and similar flow fields distort the director further, which in turn creates additional stresses and flows.

Fig.~9C is the corresponding director field, with positive and negative topological defects indicated by red and blue dots. In a passive nematic topological defects annihilate in pairs to reduce the elastic energy of the system. In an active nematic they are also created in pairs, and they often move away from each other before they can annihilate \cite{GB13,TG13,TG14,GB14}. This occurs because defects correspond to distortions in the director field and hence are sources of flow. The symmetry of the $-1/2$ defect implies balanced flows and the defect hardly moves, whereas the $+1/2$ defect moves away from the point at which it is created. Referring again to Fig.~9C note that defects are often associated with lines in the director field. These are bend walls; such a wall is shown at increased magnification in Fig.~9D. They occur because any bend instability tends to localise to give regions of nematic order separated by a narrow wall. The walls are regions of high distortion energy and therefore defects form more easily here. When the defects separate they preferentially move along the walls and in doing so they locally restore nematic order.

The dynamics of the active nematic state is summarised by the cycle in Fig.~9E. Hydrodynamic instabilities
lead to the formation of walls,  lines of high bend distortion in the director field. The elastic energy stored in the walls is released by the creation of pairs of topological defects. The director distortions that define the defects correspond to active stresses that move the $+1/2$ defect away from the $-1/2$ one. Defects preferentially move along walls, but can escape from the walls at higher activities,and when oppositely charged defects meet they annihilate. Both creation and annihilation events remove walls and help to reinstate regions of nematic order which then undergo further hydrodynamic instabilities.

\begin{figure}[htdp]
\label{corr}
\includegraphics[trim = 150 30 0 30, clip=50 200 0 200, width = 1.2\linewidth]{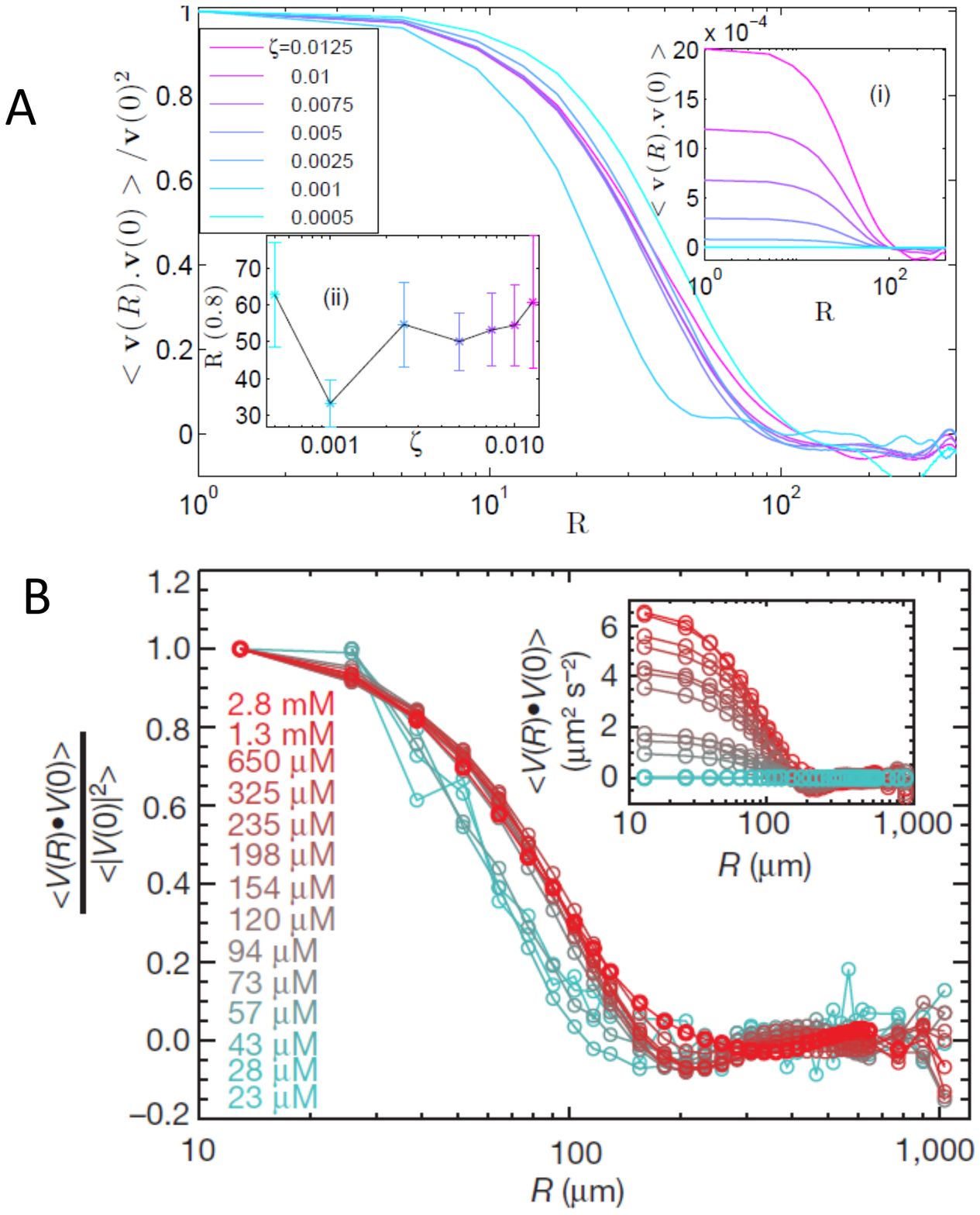}
\caption{Scaled velocity-velocity correlation function in the active turbulent state from A: simulations (after Thampi {\it et al.} \cite{TG13}), B: experiments on suspensions of microtubules and molecular motors (after  Sanchez {\it et al.} \cite{SC12}). Different colours correspond to different values of the activity (ATP concentration). The inset, top-right, is the unscaled data. The inset, bottom-left, is a measure of the vortex radius. }
\end{figure}

The velocity-velocity correlation function in the active turbulent state is plotted in Fig.~10A \cite{TG13}. The right-hand inset shows that the magnitude of the velocity increases with activity, as is expected. In the main figure the correlation functions are scaled so that their value at $R=0$ is unity. The curves collapse showing that the length scale associated with decay of correlations is independent of the activity, a feature reproduced in several experiments. 

Although I have given a description of the physical mechanisms underlying active turbulence a predictive theory is still needed. Moreover there is   still much to understand about which features of the active turbulent state are generic across different systems. For example simulations of self-propelled, non-overlapping rods with an aspect ratio $\sim 5$ show a state resembling active turbulence \cite{WD12} as do vertically vibrated granular layers \cite{NR07}. These are both systems where hydrodynamics is either absent or strongly screened.

\subsection{Microtubules and molecular motors}

A suspension of microtubules driven by molecular motors is an experimental system that displays active turbulence.\\
~\\
Molecular motors are the engines of the cell. They convert chemical energy, often in the form of ATP, into mechanical work. Kinesins are a group of motor proteins that move along microtubules through a `walking' motion of two heads which are a few nanometres in size. The heads are linked to a tether which can itself be connected to a cargo. The motors are designed to move along the microtubule tracks to overcome the strong Brownian fluctuations at these length scales. Kinesin takes an 8-nm step for every molecule of ATP it hydrolyses, and it always walks towards the head of the microtubule.

Microtubules and kinesin molecular motors form the basis of beautiful recent experiments demonstrating active turbulence \cite{SC12,HD14}.
Microtubule filaments with average length $1.5\mu$m were mixed with kinesin motors connected in clusters. The kinesin heads from a cluster form bridges between neighbouring filaments and, if the filaments have opposite polarity,  they slide relative to each other as the motors walk. This is shown in Fig.11A; note the dipolar nature of the active forces produced by the motors. The addition of PEG induces depletion interactions that result in the microtubules forming bundles and hence the effects of their relative sliding becomes much more pronounced.

\begin{figure}[htdp]
\label{motor}
\includegraphics[trim = 50 60 0 50, clip=50 200 0 100, width = 1.1\linewidth]{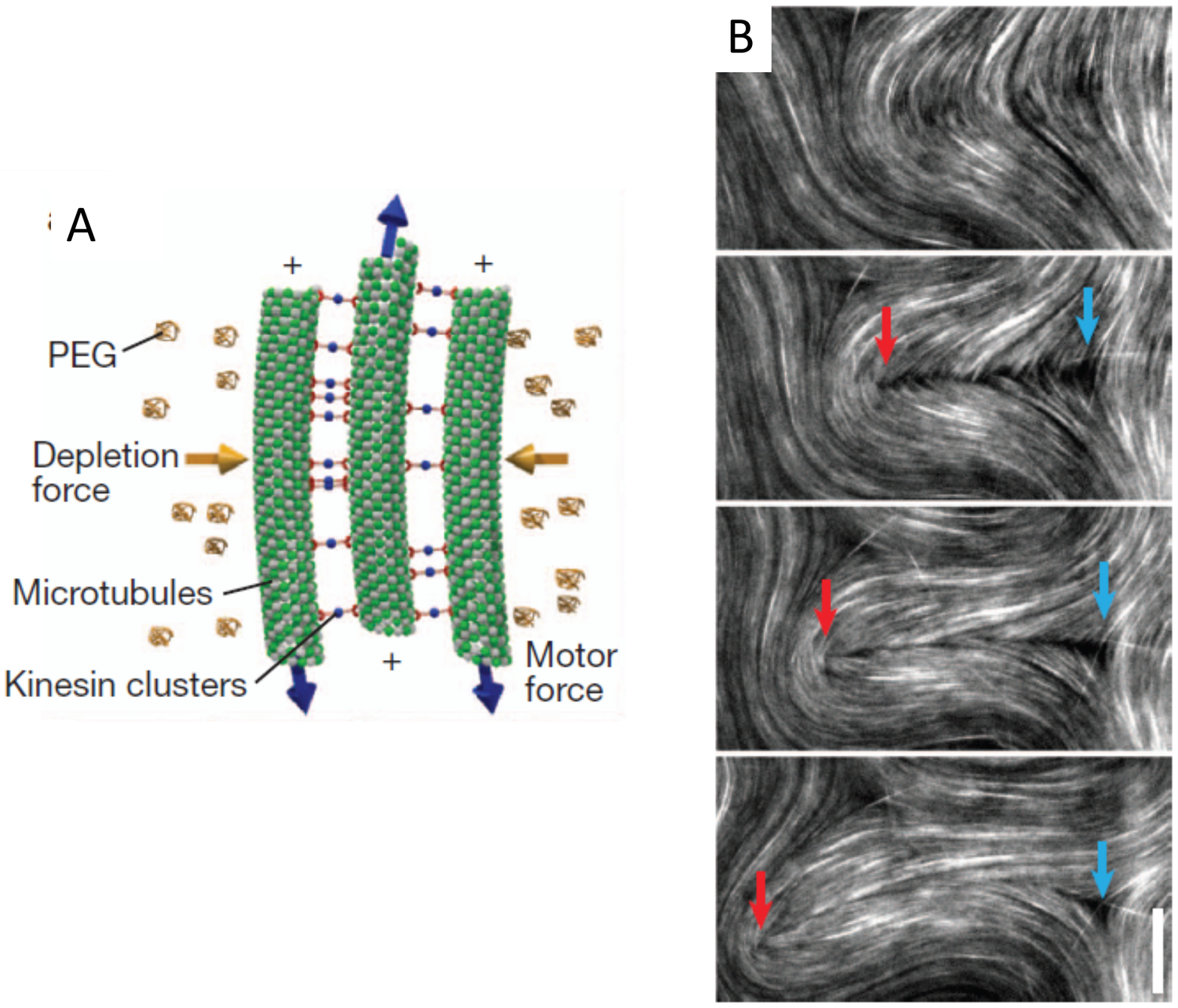}
\caption{Active turbulence in suspensions of microtubules and kinesin motors. A: The motors form bridges between the filaments which slide past each other as the motors walk. B: The microtubule bundles bend forming defects of topological charge $+1/2$ (red) and $-1/2$ (blue) (after Sanchez {\it et al.} \cite{SC12}).}
\end{figure}

A 2D active system was created by adsorbing the microtubule-kinesin mixture onto a water-oil interface. The cellular extract forms an active nematic phase characterised by swirling flows. The microtubule bundles are stretched by the motor activity and the resulting extended filaments undergo bend instabilities. This leads to the formation of $\pm 1/2$ topological defects that unbind, restoring nematic regions, as shown in Fig.~11. 

Fig.~10B shows the velocity-velocity correlation function of the microtubule-kinesin suspension for different ATP concentrations, corresponding to different strengths of the activity \cite{SC12}. Just as in the simulations the normalised correlation functions collapse onto a single curve, showing that the system has a length scale independent of the activity. However, although the creation of topological defects and the behaviour of the velocity correlation function agrees well with the model active nematic there are obvious differences between the simulations and the experiments. For example, the one elastic constant approximation should be relaxed as the microtubule bundles can be bent much more easily than splayed, the microtubules are long and flexible, and the friction of the surrounding fluid is likely to have an effect. 
Very recent work has unexpectedly shown that the $+1/2$ topological defects that are formed in the cellular suspension can themselves have nematic order which is correlated with nematic ordering of the microtubules \cite{D15}. The reason for this is currently unclear. 

In another recent paper Kleber {\it et al.} \cite{KL14}  created lipid vesicles of diameter about 70 microns and encapsulated microtubules and kinesin which absorbed onto the surface of the vesicle to form an active shell. Because of the curvature of the shell the nematic field must include defects.  
Kleber {\it et al.} observed four persistent active defects. The coupling between the velocity fields and the elastic defect-defect interactions led to the defects moving on the surface, oscillating between two symmetric tetrahedral defect states by way of a configuration where all four defects lie in a plane.  The pattern was distorted but not destroyed by thermal fluctuations. As the flexibility of the encapsulating vesicle was changed it became elliptical, fluctuating in tune with the defect oscillations. It then, on a time scale of minutes, grew four streaming, filopedia-like, protrusions, stemming from the defect sites. These reached a length of tens of micrometres  before the vesicle re-swelled and the protrusions disappeared.  


\subsection{Lyotropic active nematics}

So far we have described an active fluid at constant concentration. Adding a concentration variable to the active nematohydrodynamic equations allows us to study the coexistence between an active nematic and a passive fluid, to look at the evolution of localised patches of active material, and to introduce active anchoring.\\
~\\
We consider a binary mixture of active nematic and isotropic fluid \cite{BT14}. To distinguish the active nematic from the isotropic fluid, we define a scalar order parameter $\phi$, which measures the relative concentration of each component with $\phi=1$ for the active nematic and $\phi=0$ for the isotropic fluid. $\phi$ evolves according to the Cahn-Hilliard equation  \cite{CH58}
\begin{align}
\partial_{t}\phi+\partial_{i}(u_{i}\phi)=\Gamma_{\phi}\nabla^{2}\mu ,
\label{eqn:conc2}
\end{align}
where $\Gamma_{\phi}$ is the mobility and $\mu=\delta \mathcal{F}/\delta\phi$ is the chemical potential. A suitable choice of free energy  is
\begin{equation}
\mathcal{F} =  \frac{A_{\phi}}{2}\phi^{2}(1-\phi)^{2}+\frac{A}{2}(S^{2}\phi-\frac{1}{2}Q_{ij}Q_{ji})^{2}
+\frac{K_{\phi}}{2}(\partial_{k}\phi)^{2}+\frac{K}{2}(\partial_{k}Q_{ij})^{2}
\end{equation}
where $A_{\phi}$ and $K_{\phi}$ are material constants. An additional term $\Pi_{ij}=(\mathcal{F}-\mu\phi)\delta_{ij}-\partial_{i}\phi(\partial\mathcal{F}/\partial(\partial_{j}\phi))$ must be added to the stress components in Eq.~(\ref{elasticstress}), when the variable $\phi$ is introduced. More details of the form of the free energy and the governing equations of lyotropic active nematics can be found in  \cite{BT14}.

When a band of active nematic is placed in an isotropic fluid it is unstable to a bend instability, and undulations develop, as shown in Fig.~12A.
These are not symmetric, but have a distinctive shape with rounded convex regions, and sharper convex cusps. Topological defects of charge $+1/2$ form at the cusps, detach, and move into the bulk of the nematic, leaving the interface with a distributed negative topological charge. Note also that the director field tends to lie parallel to the interface  (Fig.~12B). There is no term in the free energy that gives a preferential anchoring; rather the anchoring, which we term active anchoring, is due to active stresses.

\begin{figure}[htdp]
\label{instability}
\includegraphics[trim = 100 10 0 30, clip=50 200 0 100, width = 1.15\linewidth]{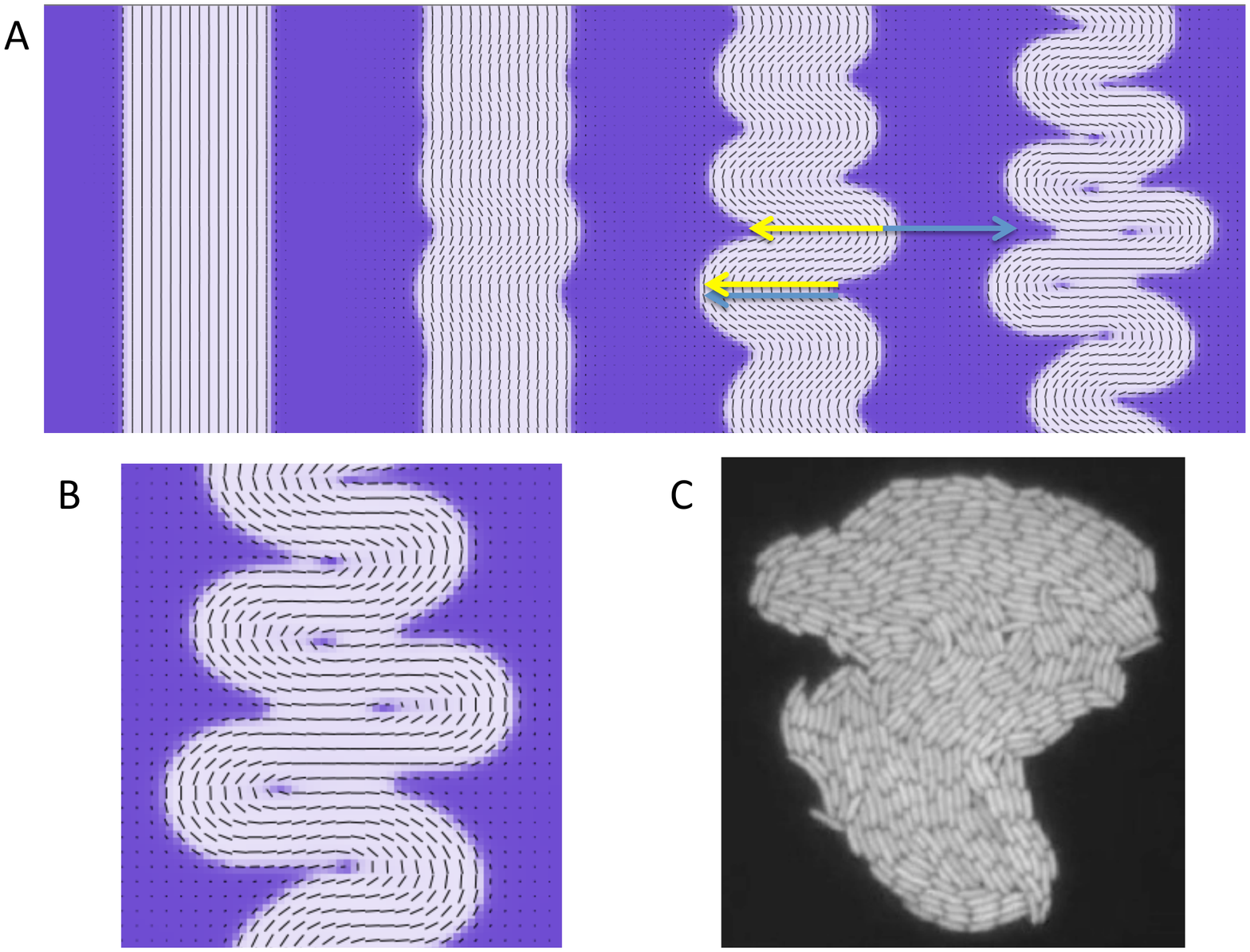}
\caption{Behaviour of an interface between an active nematic (white) and a passive fluid (purple). Short black lines represent the director field. A: Evolution with time (left to right) of an active nematic--passive interface. The interface is unstable to bend. Yellow (blue) arrows indicate the active stress induced by gradients in the magnitude (direction) of the order parameter. In convex regions these partially cancel, in concave regions they act together to give the interface its scalloped shape.
B: magnified view of the interface to illustrate planar active anchoring (after Blow {\it et al.} \cite{BT14}). C: similar planar anchoring in a dividing {\it E. coli} colony (after Stewart {\it et al.} \cite{SM05}).}
\end{figure}

To show how active anchoring arises we consider the active stresses at the interface which will result from gradients in both the magnitude and the direction of the nematic order. 
In 2D the $\bf{Q}$-tensor is 
\begin{equation}
Q_{ij} = 2q \langle   n_{i}n_{j}-\delta_{ij}/2 \rangle.
\end{equation}
We recall from Eq.~(\ref{activestress}) the active force density 
\begin{equation}
\partial_i \Pi_{ij}^{active} = - \zeta \partial_i Q_{ij}.
\end{equation}
Defining $\mathbf{m}$ as the outwards unit normal to the surface the force density can be written
\begin{equation}
-\zeta \partial_{j}Q_{ij} =-\zeta \partial_{j}q \{2 n_i n_j - \delta_{ij} \}
-\zeta q \{2n_i (\partial_j n_j) +2 (\partial_j n_i) n_j\}.
\end{equation}
If we assume that contributions from gradients in the nematic orientation are small compared to those from variations in the magnitude of the nematic order, the components of the active force density tangential and normal to the interface are
\begin{equation}
F_{\mbox{tangential}}= \zeta \mid \! \nabla q \! \mid \{2 (\mathbf{m}\cdot \mathbf{n}) (\mathbf{l}\cdot \mathbf{n})    \},
\end{equation}
\begin{equation}
F_{\mbox{normal}}= \zeta \mid \! \nabla q \! \mid \{2(\mathbf{m}\cdot \mathbf{n}) ^2- 1\},
\label{normal}
\end{equation}
where $\mathbf{l}$ is the unit vector tangent to the interface.

When the director does not lie along or perpendicular to the interface $F_{\mbox{tangential}}$ is non-zero and generates a flow along the interface. For a flow-tumbling nematic the resulting velocity gradient between the interface and the bulk nematic has a tendency to rotate the director as dictated by the corotational terms of Eq.~(\ref{eqn:cor}). In the extensile case, the director is in stable equilibrium when it lies along the interface. Conversely, in the contractile case, the flow direction is reversed, rotating the director to lie perpendicular to the interface. 

$F_{\mbox{normal}}$ acts to reinforce this effect. From Eq.~(\ref{normal}), we find that the active force density $\sim \zeta$ when $\mathbf{m}$ and $\mathbf{l}$ are parallel and $\sim -\zeta$  when they are perpendicular. Thus, for extensile activity ($\zeta>0$), the drop is extended where the interfacial alignment is parallel to the interface, and compressed where it is perpendicular to the interface, causing an initially circular drop to be stretched along the nematic director. As a result, the director field is oriented parallel to the interface everywhere except at the ends of the elongated structure. In the case of a contractile suspension, the forces are reversed, so that the nematic drop extends perpendicular to the nematic director, corresponding to active anchoring perpendicular to the interface.

To motivate the time evolution of the interface shape we now assume active anchoring and take into account gradients of $\mathbf{n}$. For the extensile system ($\mathbf{m}$, $\mathbf{n}$ perpendicular) the force normal to the interface becomes
\begin{equation}
F_{\mbox{normal}}= \zeta \{ - \!\mid \! \nabla q \! \mid -2q \;\mathbf{m}\cdot (\mathbf{n}\cdot \nabla) \mathbf{n}\}.
\end{equation}
The first term represents the force arising from the gradient in nematic order, which always acts inwards. The second term acts inwards on the concave portions of the interface, but outwards where the interface is convex. Thus, in the concave parts, the two force contributions combine to give a strong force that pulls the interface sharply inwards to give high curvature, while in the convex parts, the two contributions are opposed so the resultant force is weak leading to a lower curvature. For a contractile active nematic, the interface shape is the same because the change of sign of $\zeta$ is cancelled out by the change in active interface alignment from planar to homeotropic.

 \section{Discussion}
 
Inevitably a short lecture course contains a finite amount of material, and the choice is somewhat arbitrary.  Sections  2 and 4 of these lectures provide a general background, hopefully useful in many contexts, but many other examples could have been covered in sections 3 and 5. 
I therefore now give brief comments on some other areas that I find interesting. More details can be found in other lectures in this and related schools, and in the review articles listed.\\

\noindent
{\bf Active systems:}\\
We have considered active nematics and concentrated primarily on cases where flow provides the dominant physics. This is only one of many `active' models that have been studied since the early work of Viczek \cite{VC95} and Toner and Tu \cite{TT98}. We have not considered dry systems where any momentum correlations are short-range, the case for many of the flocking models in the literature. We have not considered polar symmetry, where additional terms appear in the continuum equations (although the active stress remains dipolar). We have ignored density variations and compressibility which are very important in the dry models, and we have neglected inertia which should be included if these ideas are to be applied to macroscopic systems such as schools of fish or animal herds. Understanding the similarities and differences between the various active models, and linking them in turn to driven systems is an important goal \cite{R10,MJ13,EW15}.\\

\noindent
{\bf Active machines and active self assembly:}\\
Much research on active matter aims to exploit or imitate the ability of microswimmers and molecular motors to act as efficient machines. We have heard in this school about active colloids which can be propelled by chemical interactions that take place over half the colloid, by anisotropic interactions induced by light, or by electric or magnetic fields \cite{EH10}. More `swimmer-like' micromachines taking the form of waving filaments and driven by an external magnetic field have also been constructed \cite{ZP10}. In an early, and appealing, example of harnessing swimmers  to do work Sokolov {\it et al.} \cite{SA10} and   Di Leonardo  {\it et al.} \cite{DA10} showed that they could drive a microscopic gear.  Colloids can be sorted into bins in an active bath of bacteria \cite{KL13} and active colloids have been shown to self-assemble \cite{SG14}.\\

\noindent
{\bf Viscocelasticity:}\\
Bacteria and molecular motors operate in  in complex environments, characterised by  confinement, crowding and strongly non-Newtonian host fluids such as mucus, extracellular matrix gels and blood. There are several results for microswimmers moving in a viscoelastic continuum \cite{L07,SL13}, but 
often the host fluid contains biopolymers or colloids of similar size to the bacteria themselves. For example, recent experiments suggest that  {\it E-coli} motility in a polymer suspension is affected by local shear-thinning induced by the flagella motion \cite{MS14}. Living liquid crystals represent a novel system where bacteria swim in a nematic fluid and their motion in the direction perpendicular to the nematic ordering is strongly suppressed \cite{ZS14}. \\

\noindent
{\bf Biological systems:}\\
Many active systems are living systems and the physical theories of active matter are starting to offer useful insight into biological mechanisms. At a sub-cellular level the active particles are molecular motors and we described in Sec 5.1 how these can drive active turbulence in dense suspensions of microtubules in vitro. Within the cell there are results showing how molecular motors walking along microtubules contribute to cell division resulting from spindle mitosis \cite{BN14}, and the interaction of myosin motors with the actin network has been well studied recently \cite{BM14}. Plant cells in particular use cytoplasmic streaming, flow driven by the motion of motors along the cell walls, presumably to aid the transport of nutrients around the cell \cite{GM15}. The extent to which hydrodynamics (even at nanometre scales) affects motor motion \cite{MP12}, the way in which multiple motors can combine to move cargo and mechanisms for cargo transport in the crowded cellular environment remain largely unexplored.

At a cellular level there is increasing evidence that cell motility is linked to the physical environment \cite{VR13}. For example, cells move up a stiffness gradient on a surface, and their rate of division is linked to the local stress. Understanding the different physical properties of cancerous cells and the link between these and the motility of malignant cells is a promising  direction \cite{GG12}. Interactions between cells, the spreading of cellular layers and the possible role of flow in morphogenesis are also of interest.\\

To summarise, in these lectures we have emphasised two important features of active swimming at low Reynolds number. Firstly at zero Reynolds number the Scallop Theorem constrains the swimmer stroke to be non-invariant under time reversal. Secondly the swimmers move autonomously and therefore can exert no net force or torque on the fluid. This means that the far flow field is generically dipolar, and its nematic symmetry has far reaching consequences for both how microswimmers stir the surrounding fluid and for their collective behaviour.

As examples of microswimmer hydrodynamics we discussed how swimmers oscillate about the centre line when swimming against a Poiseuille flow, swimmer localisation resulting from a competition between gravitaxis and shear, and the interaction of microswimmers with surfaces.  The effect of the dipolar flow field was clear when we considered the loop-like motion of tracer particles due to the flow fields produced by a passing swimmer and distinguished the contributions to tracer diffusion from tracer entrainment, swimmer re-orientations and thermal fluctuations.

In Section 4 we moved on to consider the collective behaviour of active nematics. The equations of motion are the same as those for liquid crystals but with an additional term of dipolar symmetry in the stress tensor. The active stress results in instabilities which lead to a state termed `active turbulence' characterised by high flow vorticity, and walls and topological defects in the director field. Numerical solutions to the equations of motion were compared to recent experiments on mixtures of microtubules and molecular motors. Considering an interface between an active and passive fluid, we discussed the interface instability and active anchoring, alignment of the liquid crystal at the active-passive interface that is driven by flow.

\acknowledgments

It is a pleasure to thank Gareth Alexander, Anna Balazs, Matthew Blow, Amin Doostmohammadi, J\"orn Dunkel, Ramin Golestanian,  Rodrigo Ledesma-Aguilar, Arnold Mathijssen,  Mitya Pushkin,  Victor Putz,  Miha Ravnik, Tyler Shendruk,   Henry Shum and Sumesh Thampi for many helpful and enjoyable discussions.

\end{document}